\documentclass[lettersize,journal]{IEEEtran}
\usepackage{amsmath,amsfonts}
\usepackage{algorithmic}
\usepackage{array}
\usepackage[caption=false,font=normalsize,labelfont=sf,textfont=sf]{subfig}
\usepackage{textcomp}
\usepackage{stfloats}
\usepackage{titling}
\usepackage{multirow}
\usepackage{tabularx}
\usepackage{verbatim}
\usepackage{graphicx}
\def\BibTeX{{\rm B\kern-.05em{\sc i\kern-.025em b}\kern-.08em
    T\kern-.1667em\lower.7ex\hbox{E}\kern-.125emX}}
\usepackage{balance}

\date{}
\begin{document}

\title{Parameter Efficient Fine-Tuning for \\Deep Learning-Based Full-Waveform Inversion}
\author{
    \IEEEauthorblockN{Koustav~Ghosal\IEEEauthorrefmark{1}, Abhranta~Panigrahi\IEEEauthorrefmark{2}, Arnav~Chavan\IEEEauthorrefmark{2}\IEEEauthorrefmark{4}, Arun~Singh\IEEEauthorrefmark{3},
    Deepak~Gupta\IEEEauthorrefmark{4}\\}
    \IEEEauthorblockA{\IEEEauthorrefmark{1} Indian Institute of Technology (Indian School of Mines), Dhanbad, India\\}
    \IEEEauthorblockA{\IEEEauthorrefmark{2} Nyun AI, India \\}
    \IEEEauthorblockA{\IEEEauthorrefmark{3} Department of Earth Sciences IIT Roorkee, Roorkee, India\\}
    \IEEEauthorblockA{\IEEEauthorrefmark{4} Transmute AI Lab (Texmin Hub), IIT(ISM), Dhanbad, India.}
}

%\thanks{*Corresponding Author: ads@iitism.ac.in}
% \author{
%     \IEEEauthorblockN{Koustav~Ghosal\IEEEauthorrefmark{1}, 
%     Abhranta~Panigrahi\IEEEauthorrefmark{2}, 
%     Arnav~Chavan\IEEEauthorrefmark{2}, 
%     Arun~Singh\IEEEauthorrefmark{1} and Deepak K. Gupta\IEEEauthorrefmark{2}\\}
% \IEEEauthorblockA{\IEEEauthorrefmark{1} Indian Institute of Technology (Indian School of Mines), Dhanbad, India\\}
% \IEEEauthorblockA{\IEEEauthorrefmark{2} UiT The Arctic University of Norway, Troms\o, Norway\\}
% \IEEEauthorblockA{\IEEEauthorrefmark{4} Transmute AI Lab (Texmin Hub), IIT(ISM), Dhanbad, India.}
\maketitle


\begin{abstract}
Seismic full waveform inversion (FWI) has seen promising advancements through deep learning. However, existing approaches typically focus on task-specific models trained and evaluated in isolation, leading to limited generalization across different geological scenarios. In this work, we introduce a task-agnostic foundational model for FWI, which captures general features across tasks. We first demonstrate that full fine-tuning of this foundational model outperforms task-specific models built from scratch, delivering superior performance across multiple benchmarks. Building upon this, we employ parameter-efficient fine-tuning (PEFT) to further reduce computational overhead. By fine-tuning only a small fraction of the model’s parameters, PEFT achieves comparable results to full fine-tuning, while significantly lowering memory and computational requirements. Additionally, PEFT excels in out-of-distribution tasks, where it outperforms both full fine-tuning and task-specific models. These findings establish the value of foundational modeling for FWI and highlight PEFT as an effective strategy for efficient and scalable adaptation across diverse tasks.
\end{abstract}


\section{Introduction}
% introduction to FWI and its traditional methods
Imaging the subsurface of the earth is crucial for energy exploration, carbon sequestration, reservoir identification, and earthquake warning \cite{virieux2009overview}. Among all the geophysical methods, the highest resolution image of the subsurface is obtained by using seismic full waveform inversion \cite{virieux2017introduction}.

Full waveform inversion is a technique based on full-wavefield modelling, which uses the entire content of the seismic trace to obtain a high-resolution velocity map of the subsurface. 
It is governed by the partial difference equation and performs nonlinear inversion iteratively to get the high-resolution velocity map. However, the disadvantages of physics-based FWI are high computation cost, cycle-skipping and ill-posedness \cite{deng2022openfwi}.
With the rise of deep learning (DL), data-driven methods have become an alternative to address most shortcomings of FWI. DL-based FWI studies are performed over different neural networks, like, encoder-decoder based convolutional neural network \cite{araya2018deep, wu2019inversionnet}, Recurrent Neural Network \cite{sun2020theory, sun2021physics}, Generative Adversarial Networks (GANs) \cite{zhang2020data}, etc. 

% The DL-based FWI provides an efficient alternative to yield accurate and high-quality results. However, there are still certain limitations of this strategy that limit its widespread applicability across the geoscience discipline. For example, the generalization ability of the DL-based FWI methods is limited by the availability of training data \cite{abdullin2023generalization}, and their performance on out-of-distribution cases is significantly low. Generalization is a key aspect desired from data-driven models when it comes to deployment in field (or real-world) applications. This difficulty arises from the fact that in such frameworks, the accuracy of a model's predictions heavily relies on the training data. In earth science, the distribution of one geological area can differ significantly from another, posing a strong challenge in building the right training set. Thus, minimizing generalization errors in this context is therefore non-trivial.

DL-based FWI offers an efficient alternative for generating accurate, high-quality results. However, certain limitations hinder its broader adoption in geoscience. One key challenge is the limited generalization ability of DL-based FWI methods, which is often constrained by the availability of training data \cite{abdullin2023generalization}. These models tend to perform poorly on out-of-distribution cases, which is a critical limitation when deploying them in real-world applications. Generalization is essential for data-driven models in field applications, but achieving it is difficult because the accuracy of a model’s predictions is highly dependent on the training data. In earth sciences, geological distributions can vary greatly between regions, making it particularly challenging to build an effective training set. Reducing generalization errors in this context is, therefore, a complex task.

For systematic advancements in DL-FWI, large-scale benchmark datasets have been introduced. One such benchmark is OpenFWI \cite{deng2022openfwi}, a comprehensive, multi-structural dataset spanning various geological features, including interfaces, faults, and field data. OpenFWI provides diverse 2D and 3D datasets for training and evaluating deep learning models, encouraging generalization across different geological scenarios. Among the models benchmarked within OpenFWI, InversionNet has emerged as an effective baseline model for seismic inversion tasks, demonstrating strong performance on specific target datasets.

Despite its success in individual tasks, the task-specific models built using InversionNet, often struggle to generalize across different tasks and geological variations, as noted in \cite{wu2019inversionnet, lozenski2024learned}. This limitation highlights the need for models that can maintain robust performance across a wide range of geophysical scenarios, an open research direction that demands further exploration.

To address the challenges of generalization, recent advancements in deep learning have introduced Pretrained Foundational Models (PFMs) \cite{zhang2024improving, boiarsky2023deep}. PFMs leverage large, diverse datasets to build a robust base model that can be fine-tuned for various downstream tasks. These models have proven highly effective in fields like natural language processing and computer vision and are now making their way into geophysics. For example, the Seismic Foundation Model (SFM) \cite{sheng2023seismic} represents an initial effort to create PFMs specifically for seismic applications, aiming to provide a unified foundation for different geophysical tasks.

However, the direct application of PFMs to geophysical tasks like FWI is challenging due to the statistical variations in data acquired through different geophysical methods, differences in spatio-temporal resolution, and the distinct physical meanings of the data \cite{sheng2023seismic}. Fine-tuning remains essential to adapt PFMs effectively to downstream tasks. However, traditional full fine-tuning, which involves updating all model parameters, is computationally intensive and can lead to overfitting, especially in low-data scenarios.

To mitigate these issues, recent research has introduced Parameter-Efficient Fine-Tuning (PEFT) methods \cite{ding2023parameter}. PEFT techniques, such as Low-Rank Adaptation (LoRA) \cite{hu2021lora}, enable adaptation to new tasks by updating only a subset of model parameters or by using lightweight adapters, significantly reducing computational and memory costs while maintaining performance. These approaches hold significant promise for DL-FWI, enabling efficient model adaptation and improved generalization, even in low-data regimes.

The focus of this paper is to explore how PEFT, and by extension, pretrained foundational models, can be leveraged to develop models that generalize effectively across diverse deep learning tasks, particularly in the context of FWI. By examining the integration of PEFT with PFMs, we aim to provide insights into building more adaptable, task-agnostic models capable of robust performance across varied geological settings.

The contributions of this paper can be summarized as:
\begin{itemize}
\item We propose the concept of building foundational models that can handle different geological features, demonstrating that this approach achieves better performance compared to task-specific models in FWI.

\item We demonstrate how PEFT, specifically through the LoRA method, offers an efficient and effective solution for fine-tuning pretrained models. LoRA performs at par with full fine-tuning in in-distribution (ID) and outperforms out-of-distribution (OOD) samples, demonstrating improvements in low-data scenarios.

\item By integrating PEFT with foundational models, we achieve enhanced generalization across tasks, improved performance in low-data regimes, and reduced memory consumption, making this approach highly suitable for DL-FWI applications.

\end{itemize}

\section{Background}
\label{sec: background}
\subsection{Seismic full waveform inversion}

Full Waveform Inversion (FWI) is a seismic imaging technique used to obtain high-resolution subsurface velocity maps. It is based on full-wavefield modelling, utilizing the complete seismic wavefield recorded at the surface to iterative invert for subsurface properties. Unlike conventional seismic methods, which rely on only part of the wavefield (e.g., first arrivals or reflected waves), FWI makes use of the entire content of the seismic trace, capturing both amplitude and phase information to enhance imaging accuracy.

FWI is governed by partial differential equations that describe wave propagation, typically the acoustic wave equation for isotropic media. This technique involves nonlinear inversion, where synthetic wavefields are generated and iteratively matched to observed wavefields to minimize the difference between them. Despite its ability to produce detailed subsurface velocity models, traditional physics-based FWI faces significant challenges, including high computational cost, cycle-skipping, and ill-posedness, especially in complex geological settings.

\textbf{Acoustic Wave Equation in FWI. }In this paper, we focus on acoustic FWI, which is widely used in practice due to its relative simplicity and computational efficiency. In an isotropic medium with constant velocity, forward modelling in FWI is described by the acoustic wave equation:

\[
\nabla^{2}p - \frac{1}{c^2}\frac{\partial^2 p}{\partial t^2} = s
\]

Here, \( p \) represents the pressure field, \( s \) is the seismic source term, and \( c \) is the velocity of the subsurface, which varies spatially as \( c = c(x,y,z) \). The pressure field \( p \) is a function of both spatial coordinates and time, denoted as \( p = p(x,y,z,t) \). The Laplacian operator \( \nabla^2 \) is defined as:

\[
\nabla^2 = \frac{\partial^2}{\partial x^2} + \frac{\partial^2}{\partial y^2} + \frac{\partial^2}{\partial z^2}
\]

In forward modelling, given the source term \( s \) and the velocity map \( c \), the pressure field \( p \) is computed by solving the acoustic wave equation, representing the wavefield at the receiver location. This forward modelling process can be conceptualized as a function \( f(c) \), where the velocity map \( c \) is mapped to the pressure field \( p \).

\textbf{Inverse Problem in FWI. }The core objective of FWI is to invert the observed pressure field \( p \) to estimate the subsurface velocity map \( c \). Mathematically, this is expressed as the inverse mapping:

\[
c = f^{-1}(p)
\]

Where \( f^{-1} \) represents the inversion of the forward modelling function. In the conventional FWI workflow, forward modelling is used to generate synthetic data, which is then compared to the true observed data. The difference between the two is measured by a cost function, and the gradient of this cost function with respect to the velocity model parameters is calculated to update the velocity map. This iterative process continues until the synthetic data sufficiently matches the true data.

\subsection{Parameter-Efficient Fine-Tuning (PEFT)}
\label{sec:peft}
% Q.> Why PEFT?
Large-scale deep learning models exhibit a remarkable degree of generalization ability, enabling them to transfer their knowledge to novel tasks that were entirely unknown during the training process. Due to the large scale of the models, it is not recommended to fully fine-tune all the parameters of the model. PEFT involves selectively adjusting a small number of parameters in a pre-trained model while leaving the others unchanged. In this way, the model can adapt to different downstream tasks with less computation and few label data \cite{han2024parameter}.

% Different types of PEFT
Broadly PEFT can be classified into three categories:
\begin{enumerate}
    \item \textbf{Additive method}: Unchanged pre-trained backbone and only a few trainable parameters are introduced. Due to the addition of a few trainable parameters, this method is known as the additive method. The popular methods that come under additive methods are:
    \begin{enumerate}
        \item \textit{Task-Specific Adapters}: The task-specific classifier $f_{(\theta, \phi)}$ was developed from learned task-specific weights $\phi$. Network minimizes the loss over the downstream task w.r.t task-specific weights $\phi$.
        
        \item \textit{Scale-Shift Features}: Scale ($\gamma$) and shift ($\beta$) are two parameters that are used for feature modulation. The output from the previous operation is scaled by the $\gamma$ via a dot product and then adjusted by adding the $\beta$.
        $ y = \gamma * x + \beta$  
        
        \item \textit{LoRA}: The idea is that when adapting, the weight updates show a low inherent rank.
        \begin{equation}
            W + \Delta W = W_0 + \alpha BA
        \end{equation}
        where pre-trained weight matrix $W_0$, $\Delta W$ is the weight update that is defined by the low-rank decomposition with rank r, $B \in \mathbb{R}^{d \times r} \textsc{ and } A \in \mathbb{R}^{r \times k}$. LoRA 
        Alpha ($\alpha$) is the scaling factor, a higher value of alpha gives more weight to the low-rank structure whereas, a lower value of alpha reduces the influences over the low-rank adapters and makes the model more dependent on the original pretrained model
        \label{lora}
        \item \textit{AdaptFormer}: AdaptFormer module (AdaptMLP) contains two parallel branches. The first branch is the MLP block of the vanilla transformer and the second branch includes a down-projection $W_{down}$, a ReLU activation layer, an up-projection $W_{up}$ and a scaling factor (s). The modified features are merged with the original input features entering the AdaptMLP block via a residual connection.
    \end{enumerate}
    \item \textbf{Selective method}: Instead of adding more parameters which increases the model complexity. It selects a subset of the existing parameters used to fine-tune for the downstream task. Some of the popular methods for the selective methods are:
    \begin{enumerate}
        \item BatchNorm Tuning
        \item Bias Tuning
        \item Attention Tuning
        \item LayerNorm Tuning
        \item BitFit
    \end{enumerate}
    \item \textbf{Prompt tuning}: Visual prompt tuning enhances Transformer models by wrapping the original input with trainable prompts, which are optimized for specific tasks to align the input distribution with pre-training data.
\end{enumerate}

\section{Proposed Methodology}
% \subsection{Problem Statement}
% Let $ D=(x_i,y_i)_i^N $ be the downstream velocity dataset of interest, which consists of inputs seismic data $x_i$ and their corresponding velocity map $y_i$. Let $f$ be a pre-trained model parameterized by $\theta$. Let $\mathcal{L}$ be a loss function we want to minimize. Our objective is to minimize the total loss $L = \frac{1}{N}\sum_i^N\mathcal{L}(f(x_i;\theta),y_i)$ using gradient descent, starting with the initialization $\theta = \theta_0$, where $\theta_0$ are the weights from pre-training. Such thorough fine-tuning isn't always feasible due to resource limitations. Since several layers may have learnt broadly relevant characteristics, it may not be desirable to adjust all of the network weights. Additionally, the limitation in the availability of the velocity dataset plays another major role in limiting the performance of the downstream tasks.

\subsection{General description}

\begin{figure*}[!h]
    \centering
    \includegraphics[width=0.75\textwidth]{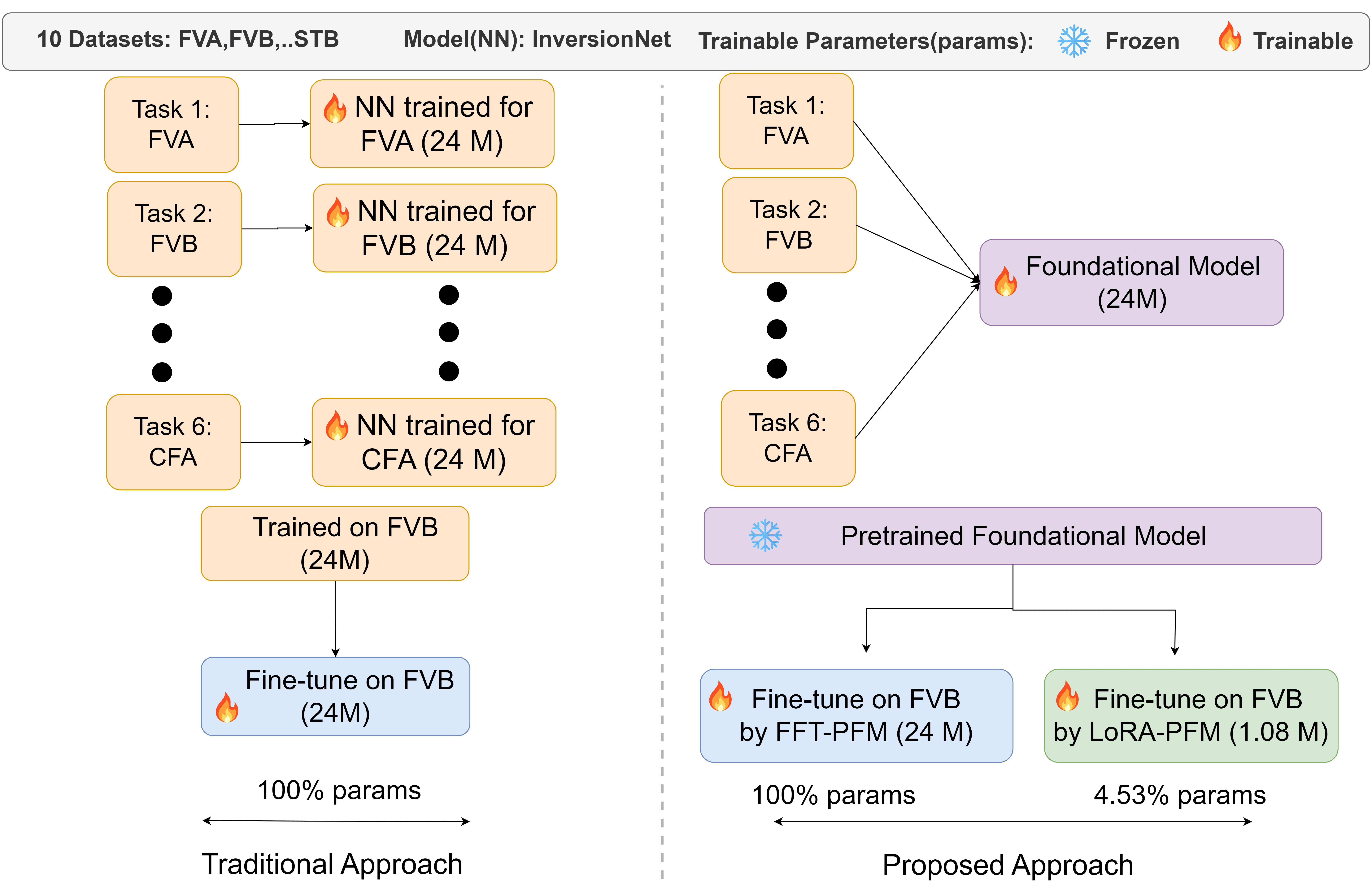}
    % \caption{ Comparison of our proposed approach with the traditional fine-tuning method. FVA: FlatVel A, FVB: FlatVel B, CFA: CurveFault A, CFB: CurveFault B, STB: Style B.
    % Proposed Approach }
    \caption{Comparison of our proposed approach with the traditional fine-tuning method. The left-hand side figure shows the traditional task-specific model and traditional fine-tuning method. The right-hand side figure represents the pertaining of the foundational model and subsequent two types of fine-tuning methods: Full fine-tuning(FFT-PFM) and LoRA-PFM.}
    \label{fig:proposed_approach}
\end{figure*}

The conventional approach in DL for geophysical applications relies heavily on task-specific models, which are designed and trained for a particular dataset or task using a significant amount of labelled data. However, generating large labelled geophysical datasets is not only labour-intensive but also computationally demanding. For instance, creating a dataset of one million geophysical data pairs (e.g., seismic data and corresponding velocity models) requires running one million forward simulations, which is time-consuming. As a result, these task-specific models often struggle with data scarcity, leading to overfitting and poor generalization to new or unseen geological scenarios.

Fig. \ref{fig:proposed_approach}(a) illustrates this traditional task-specific modelling approach, where models like InversionNet are trained and tested on the same dataset, such as FlatVel-B from OpenFWI. This setup emphasizes that the model's ability to generalize is inherently limited to the characteristics of the dataset it was trained on, making it less effective when applied to different geological features or OOD tasks.

In response to these limitations, we propose a novel approach that leverages PFMs combined with Parameter-Efficient Fine-Tuning (PEFT). 
PFMs are large-scale models initially trained on a broad, diverse dataset, allowing them to capture rich representations that can generalize across a wide range of geophysical tasks. Unlike task-specific models, PFMs are designed to serve as a base model that can be efficiently adapted to various downstream tasks with minimal labelled data. This flexibility enables better performance, even in data-scarce scenarios, while also reducing computational costs during adaptation.

The proposed methodology consists of two main stages, as depicted in Fig. \ref{fig:proposed_approach}(b):

\textbf{Pretraining the Foundational Model:} In the first stage, we build a foundational model by pretraining it on a large, diverse dataset from OpenFWI, encompassing various geological features. This extensive pretraining phase allows the model to learn generalized representations of seismic data and velocity maps, forming a robust base for subsequent adaptations.

\textbf{Task Adaptation Using PEFT}: In the second stage, we use PEFT methods to adapt the foundational model to different downstream tasks, such as FlatVel B, CurveFault B, Style A, and Style B. PEFT techniques allow us to modify only a small subset of parameters, preserving the core knowledge captured by the foundational model while enabling efficient adaptation to new geological scenarios. This step is crucial for enhancing performance in both in-distribution (ID) and OOD samples, as it ensures that the model retains its generalization capabilities while fine-tuning to specific tasks.

\subsection{Building a task-agnostic foundational model}
\label{pretrain_setup}
% why do we have build pfm?
% The concept of foundational models in deep learning refers to a new class of large-scale, pre-trained models that serve as a base for solving a variety of downstream tasks across different domains. These models are first trained on massive datasets, using supervised learning techniques, where they learn to understand general patterns and representations from raw data and labels. 
%PTMs offer a more efficient alternative compared to the traditional deep learning approach, which depends heavily on large-scale, task-specific, and crowd-labeled data to train individual deep neural networks (DNNs) for different visual recognition tasks \cite{azad2023foundational}. They are pretrained on a large dataset, which improves the accuracy of the model in case of limited downstream tasks. The limited data availability reduces the generalization error, whereas a significantly large training dataset improves the generalization of PFMs. Such a shift in approach results in a significant reduction in labour and time. 

% In the above section, we have seen that PTMs introduce a paradigm shift from the traditional DL. In transfer learning, two pre-training strategies are often investigated extensively: feature transfer and parameter transfer. We have used supervised pretraining with parameter transfer, in which we train the parameter with the source task and fine-tune the target task with the shared parameters. 

In the previous section, we discussed how PFMs represent a paradigm shift from traditional DL methods, which was driven by advances in transfer learning and large-scale neural networks \cite{bommasani2021opportunities}. 
%These models are pre-trained on massive datasets, capturing generalizable features that can be transferred to a variety of tasks, thus enabling flexible application across domains. 
Transfer learning plays a crucial role in the success of PFMs, with two primary approaches explored: feature transfer and parameter transfer. For this study, we employed supervised pretraining using the parameter transfer method, where the model is first trained on a source task and then fine-tuned on a target task by leveraging the shared parameters learned during pretraining. %Large-scale networks are critical for building foundational models because they enable the models to learn rich, generalized representations from vast amounts of data, which is essential for transferability across diverse tasks. The neural network used for pretraining is InversionNet, which is based on a U-shaped encoder-decoder (U-Net) architecture. The model's scale can be increased by expanding both the encoder and decoder paths, allowing it to adapt to more complex tasks. %Additionally, foundational models benefit from their ability to handle increasingly large datasets and more sophisticated tasks, providing the flexibility to apply pre-trained knowledge to new domains with minimal retraining effort.
For this study, we employed InversionNet, an encoder-decoder network based on a U-shaped architecture (U-Net), as the backbone for our PFM. InversionNet's architecture consists of several convolutional blocks in the encoder and decoder paths, connected via skip connections. Each convolutional block includes a convolutional layer, batch normalization, and a leaky ReLU activation layer. The encoder compresses the input data from a size of bs$\times$5$\times$1000$\times$70 to bs$\times$512$\times$1$\times$1, where bs is the batch size (bs = 256). The decoder then reconstructs the data to its original dimensions using transposed convolutional layers. This architecture allows InversionNet to learn complex features from seismic data, making it an ideal candidate for building a robust foundational model.

To create our PFM, we trained InversionNet using the OpenFWI dataset, a comprehensive collection of large-scale, multi-structural benchmark datasets designed for full waveform inversion (FWI). OpenFWI is categorized into four groups: the "Vel Family," "Fault Family," "Style Family," and "Kimberlina Family." For this study, we focused on the two-dimensional datasets from the "Vel Family," "Fault Family," and "Style Family," excluding the three-dimensional "Kimberlina Family".
The datasets are further divided based on subsurface complexity into easy (A) and hard (B) versions. The "Vel Family" and "Fault Family" are also categorized into flat (Flat-) and curved (Curve-) subsurface shapes. The details of the datasets used for pretraining are summarized in Table I. The seismic data and velocity maps have dimensions of 5$\times$1000$\times$70 and 70$\times$70, respectively. We used supervised pretraining with a total of 216,000 training samples and 36,000 test samples.
Pretraining InversionNet on this diverse dataset allows the model to learn generalized features from various geological structures, enhancing its ability to handle complex tasks and improving its robustness in real-world applications. This foundational knowledge significantly reduces the task-specific data and time needed for fine-tuning, making the model more adaptable and efficient for different FWI tasks. 
% Since PFMs rely on large datasets, we used a publicly available dataset called OpenFWI. This dataset provides a wide range of large-scale, multi-structural benchmark datasets, which are categorized into four groups: the "Vel Family," "Fault Family," "Style Family," and "Kimberlina Family." We excluded the "Kimberlina Family" because it contains three-dimensional data, as this study focuses only on two-dimensional FWI problems.

% The three families are split into two groups based on subsurface complexity: an easy version (A) and a hard version (B). The "Vel Family" and "Fault Family" are further divided into flat (Flat-) and curved (Curve-) categories, depending on the shape of the subsurface. The details of the dataset are given in the table.
% \ref{tab:pretrain_data_details}.
\begin{table}[!h]
    \centering
    \caption{\MakeUppercase{Pretraining dataset details: collection of six dataset from OpenFWI \cite{deng2022openfwi} }}
    \begin{tabular}{|m{6em}|m{7em}|m{2em}|m{3.5em}|m{3em}| }
    \hline
    Group & Dataset & Size & Train set & Test set  \\
    \hline
    \multirow{2}{6em}{Velocity Family} & FlatVel-A & 43G & 24K & 6K \\
    & CurveVel-A/B & 43G & 24K & 6K \\
    \hline
    \multirow{2}{6em}{Fault Family} & FlatFault-A/B & 77G & 48K & 6K \\
    & CurveFault-A & 77G & 48K & 6K \\
    \hline
    \end{tabular}
    \label{tab:pretrain_data_details}
\end{table}
% describe the process of building PFM.
% The size of the seismic data and velocity map are 5$\times$1000$\times$70 and 70$\times$70 respectively. We have used supervised pretraining with the total number of train and test samples for InversionNet as 216,000 and 36,000, respectively. 
% Fig. \ref{fig:proposed_approach}(b) shows the schematic view of the proposed approach for pretraining and fine-tuning to different datasets. Pretraining with a diverse dataset on a large model like InversionNet offers key benefits for FWI tasks, including improved generalization by learning features from various geological structures, which enhances performance on unseen data. It allows the model to handle complexity by being exposed to both easy and hard dataset versions, thereby improving its robustness in real-world applications. Additionally, pretraining provides foundational seismic knowledge that reduces the amount of task-specific data and time needed for fine-tuning.

\subsection{Building task-specific models with PEFT}
% \dpk{Describe why and how the pre-trained foundational model is adapted to every task.}
% Q> why?
% PEFT methods have emerged as a crucial approach in the context of large-scale pre-trained models due to their ability to address several significant challenges.  Full fine-tuning of large models, such as BERT \cite{devlin2018bert}, GPT, or T5, requires updating millions or even billions of parameters, which demands substantial computational resources, memory, and storage. PEFT methods mitigate this issue by only updating a small subset of model parameters, resulting in reduced memory and storage requirements, as well as faster training times \cite{houlsby2019parameter}. Detailed discussion on PEFT was shown in section \ref{sec:peft}. These methods are particularly advantageous in scenarios involving multiple tasks, as fully fine-tuning large models for each new task is inefficient. 
In this study, we define tasks as recovering velocity structures for specific geological scenarios, such as recovering fault structures (CurveFault), layered structures (FlatVel), and more complex, randomly distributed diverse structures (Style). Apart from the various PEFT methods, we have also utilized LoRA to fine-tune our pretrained model to different velocity datasets (or tasks).
The operational aspects of LoRA are detailed in section \ref{sec:peft} \ref{lora}. LoRA was created for Transformer Language Models, but it can also be utilized with any neural network that has dense layers \cite{hu2021lora}. 
For this study, we have used two fine-tuning strategies: full fine-tuning of the PFM (FFT-PFM) and low rank adaptation of PFM (LoRA-PFM). The key advantages of LoRA-PFM over FFT-PFM are listed below:
\begin{enumerate}
    \item LoRA-PFM reduces computational cost by learning low-rank updates instead of full fine-tuning of all parameters(i.e., FFT-PFM), enabling faster and cheaper training \cite{hu2021lora}.
    \item LoRA-PFM preserves the pretrained model’s original weights, mitigating catastrophic forgetting and allowing reuse of the same model across multiple tasks with lightweight task-specific updates.
    \item LoRA-PFM is highly efficient in switching between various tasks by simply replacing the LoRA adapters (1.1 M), while the same with FFT-PFM becomes infeasible due to increasing storage demands (24 M)  \cite{chung2024scaling}.
\end{enumerate}
We applied LoRA-PFM to InversionNet, a CNN-based network. Implementing LoRA in InversionNet requires adding LoRA layers to both convolutional and transposed convolutional layers. Each LoRA layer (or LoRA module), employs two trainable matrices: $A$ and $B$. The LoRA matrices (LoRA $A$ and LoRA $B$) provide task-specific adjustments by projecting the input into a lower-dimensional space with $A$, and then reverting it to the original space using $B$. A scaling factor ($\alpha$) is used on the output of the LoRA layer to manage the size of the LoRA rank updates.
% The working details of LoRA-PFM are explained in the section \ref{sec:peft} 1) c).
% LoRA-PFM was developed for Transformer Language Models but could be applied to any neural network with dense layers \cite{hu2021lora}. In this study, we have used LoRA-PFM on a CNN-based network, implementing LoRA-PFM in the InversionNet model involves introducing trainable low-rank matrices into the convolutional layers and transpose-convolutional layers to fine-tune the model efficiently. LoRA-PFM introduces two trainable matrices: LoRA-PFM A and LoRA-PFM B. The LoRA-PFM matrices (LoRA-PFM A and LoRA-PFM B) introduce task-specific updates by projecting the input into a low-dimensional space using LoRA-PFM A and then projecting it back to the original space using LoRA-PFM B. A scaling factor($\alpha$) is applied to the output of the LoRA-PFM layer to control the magnitude of the LoRA-PFM rank updated.
% In this study, we have used LoRA-PFM to CNN-based model. InversionNet is a CNN-based encoder-decoder network with convolutional and transpose convolutional layers. LoRA-PFM layers are added to convolutional as well as transpose-convolutional layers. 
LoRA-PFM modules are trained for each dataset (CurveFault B, FlatVel B, Style A, and Style B) and these small trained modules are stored rather than duplicating the entire model for each task or dataset. This flexibility allows the same pre-trained model to be adapted to various tasks by swapping LoRA-PFM modules, enabling efficient use of memory and storage resources.
Our experiments show that LoRA-PFM is an efficient alternative to the fine-tuning method in terms of accuracy, memory and generalization. 

%These facts are described in the experiment section, which is further divided into various subsections. The subsection \ref{sec: pretraining} describes the pretraining process. The performance of various methods in ID and OOD scenarios is discussed in subsection \ref{sec: generalization}. The subsection \ref{sec: low data} discusses the performance of the models in the presence of low data. The final subsection \ref{sec: memory} discusses the memory efficiency of the proposed method.

\section{Experiments}

\subsection{Implementation details}
The implementation details regarding the hyperparameters, loss function and evaluation metrics are summarized in the supplementary material. We consider the L1-norm as the loss function to train the PFM and also in finetuning. Three evaluations used for this study are: MAE, RMSE, and SSIM.
\subsection{Performance evaluation of the foundational model}
\begin{figure}
    \centering
    \includegraphics[width=\linewidth]{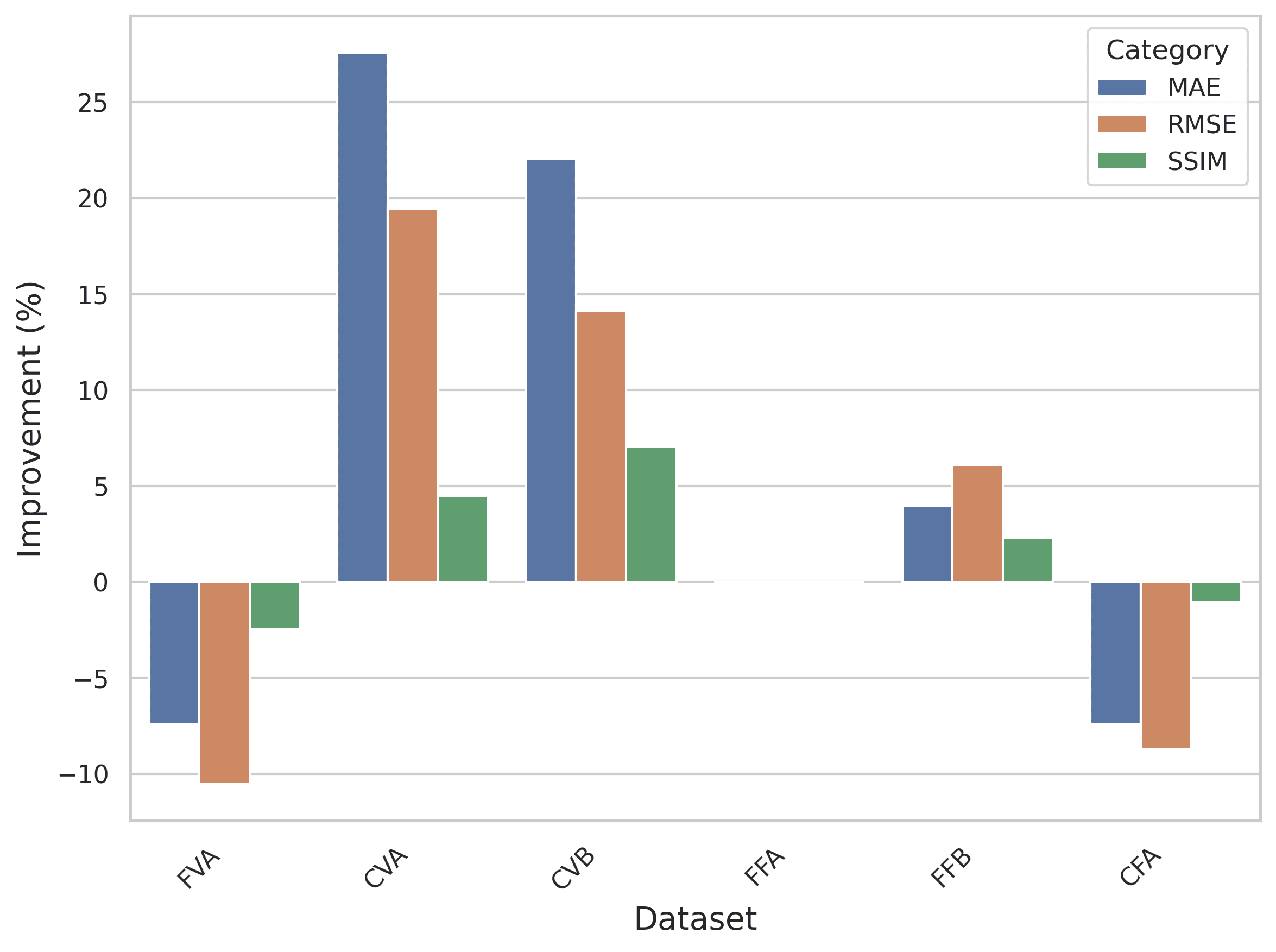}
    %\caption{Improvement in performance of the Pretrained Foundational Model compared to the baseline model.}
    \caption{Performance improvement of the pretrained foundational model trained on six datasets from OpenFWI is analyzed over the baseline in terms of three accuracy metrics: MAE, RMSE, and SSIM. The abbreviation are FVA: FlatVel A, CVA: CurveVel A, CVB: CurveVel B, FFA: FlatFault A, FFB: FlatFault B, CFA: CurveFault A}
    \label{fig:pretrain_baseline_table}
\end{figure}

% \textbf{Result:}\\
The design of the pretrained foundational model (PFM) is discussed in section \ref{pretrain_setup}. In this section, we will analyse the results of the pretraining on six 2D OpenFWI datasets. We evaluate the performance metrics and compare them against the baseline model to determine the effectiveness of our proposed approach. The results demonstrated in Fig. \ref{fig:pretrain_baseline_table} indicate the PFM demonstrates significant improvement in the complex datasets, such as CurveVel A/B, and FlatFault B and closely follows the baseline model in simple datasets, such as FlatVel A, FlatFault A and CurveFault A. The complex datasets have high spatial information, here we refer to spatial information \cite{yu2013image} as the mean of gradient magnitude on horizontal and vertical direction via the Sobel filter. %The Sobel filter is an edge detection operator that calculates the gradient of the image intensity function. It uses convolution with two 3x3 kernels, one for detecting changes in the horizontal direction (Gx) and the other for the vertical direction (Gy). 
% \begin{equation}
%     SI_{mean} = \mathop{\mathbb{E}}\sqrt{(G^2_x + G^2_y)}
% \end{equation}
% *********** old***************
% In the case of CurveVel A, we observe a significant improvement: the Mean Absolute Error (MAE) score decreases from 0.066 to 0.050 from the baseline to the pre-trained model (PI). The Root Mean Square Error (RMSE) decreases from 0.124 to 0.102, and the Structural Similarity Index (SSIM) increases from 0.812 to 0.849.
% Similar performance gains are observed for CurveVel B and FlatFault B. However, for other datasets with less spatial information, the scores are lower than the baseline. This suggests that pretraining helps the network perform better with more complex datasets rather than simpler ones. 
% **************** New *******************
% In Fig. \ref{fig:pretrain_baseline_table}, we observe a significant improvement for CurveVel A, CurveVel B, and FlatFault B in terms of MAE, RMSE, and SSIM. However, for FlatVel A and CurveFault A, there is a decline in the performance of PFM across all accuracy metrics. Additionally, for FlatFault A, no significant improvement is achieved compared to our baseline.
% From the above discussion, we observe that pretraining significantly improves over more challenging datasets like CurveVel A, CurveVel B, and FlatFault B in comparison to simple datasets such as FlatVel A, FlatFault A, and CurveFault A which shows minimal improvement. 
However, for simpler datasets such as FlatVel A, FlatFault A and CurveFault A, we see a small dip in performance and relatively better performance in complex datasets such as CurveVel A, CurveVel B, and FlatFault B.
This is primarily due to the focus of our PFM towards generalizing on a broader set of features rather than fitting towards a narrower set. We believe that additional training might be able to eliminate this gap as well. Nevertheless, it is clearly evident that PFM on average improves performance over the baseline model.
The qualitative analysis is provided in Section II of the supplementary material. This suggests that pretraining results in enhanced feature extraction, which allows the model to capture intricate patterns within the data.

\subsection{Fine-tuning of the foundational model}
Fine-tuning foundational models usually involves updating all model parameters to adapt the PFM for a specific downstream task. During this process, the PFM is fine-tuned with a smaller, task-specific dataset and uses the L1-norm as the loss function. This approach allows the model to retrain its general knowledge while also capturing specific patterns and features relevant to the new task. Despite its effectiveness, conventional fine-tuning is computationally expensive and memory-intensive, particularly for large foundational models with billions of parameters. Furthermore, it poses a risk of overfitting when applied to small datasets, as updating all parameters can lead to the degradation of the general features learned during pretraining. This approach also requires storing a separate, full copy of the fine-tuned model for each task, making it inefficient in scenarios requiring multi-task learning or task-switching.

\begin{figure}
    \centering
    \includegraphics[width=0.9\linewidth]{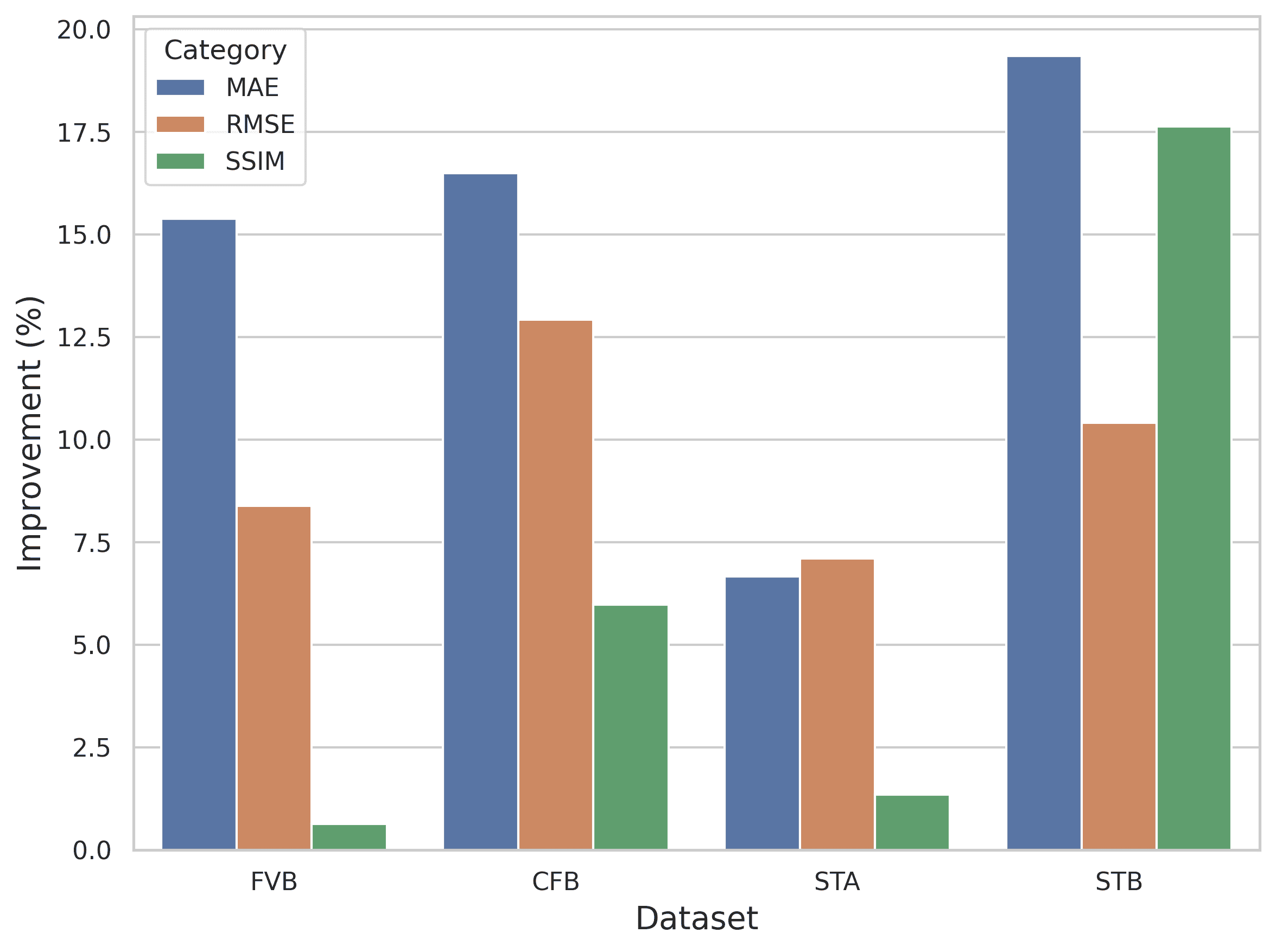}
    \caption{Performance improvement of the PFM, when fine-tuning on four datasets over the baseline in terms of MAE, RMSE, and SSIM.}
    \label{fig:fine-tune pretrained model}
\end{figure}

\textbf{Results: }In Fig \ref{fig:fine-tune pretrained model}, the statistical performance of PFM over four unseen datasets from OpenFWI demonstrates significant improvements over task-specific training (baseline).
% The detailed analysis shown in the supplementary table \ref{tab:finetune_pretrain_model} reveals that the pretrained model outperforms the baseline across all datasets.
The statistical analysis shows that FFT-PFM performs better than the baseline across all four datasets.
For FlatVel B, we can observe an improvement of 15.3\% in MAE, 8.3 \% in RMSE and 0.63 \% in SSIM. These improvement percentages are shown in Fig. \ref{fig:fine-tune pretrained model} and represent the performance improvement of FFT-PFM in comparison to the baseline. From Figure  \ref{fig:fine-tune pretrained model}, we can observe that the maximum improvement occurs for CurveFault B and Style B and these two datasets are the most complex within the group of four. So, we can observe that the FFT-PFM improves more over the complex dataset with respect to the simple dataset.  
%Improvement percentages show that FFT-PFM significantly improves the accuracy of the predicted velocity data compared to the baseline model. 
Apart from the MAE and RMSE, we can observe that SSIM shows a minimal improvement for FlatVel B, and Style A, indicating that the structural similarity stays mostly the same in both methods. 
For CurveFault B, we also observe a similar pattern of improvement. However, here SSIM is significantly large, indicating an enhancement in the accuracy of the predicted velocity structures as well as the velocity values. Improvement in Style A is small compared to all the other datasets since Style A is a simple dataset. Both the methods' perform well but FFT-PFM is slightly above the Baseline. For Style B, being the most complex dataset, improvement is most pronounced. The baseline fails to capture the intricate details and the difference between FFT-PFM and baseline becomes large. The qualitative results that support the statistical analysis are presented in Section III of the supplementary document.
These enhancements demonstrate a more accurate alignment between the predicted velocity map and the ground truth, surpassing the results achieved by the baseline method.

% \begin{table}[!h]
%     \centering 
%     \begin{tabular}{|c|c|c|c|c|c|}
%     \hline
%     Data & Method
%     & MAE ($\downarrow$) & RMSE ($\downarrow$) & SSIM ($\uparrow$) \\
%     \hline
%     \multirow{2}{6em}{FlatVel\_B} & Baseline & 0.035 & 0.087 & 0.946 \\
%     & PT & \textbf{0.030} & \textbf{0.080} & \textbf{0.952} \\
%     \hline
%     \multirow{2}{6em}{CurveFault\_B} & Baseline & 0.164 & 0.247 & 0.616 \\
%     & PT & \textbf{0.139} & \textbf{0.217} & \textbf{0.654} \\
%     \hline
%     \multirow{2}{6em}{Style\_A} & Baseline & 0.062 & 0.102 & 0.885 \\
%     & PT & \textbf{0.058} & \textbf{0.095} & \textbf{0.897} \\
%     \hline
%     \multirow{2}{6em}{Style\_B} & Baseline & 0.068 & 0.091 & 0.631 \\
%     & PT & \textbf{0.056} & \textbf{0.082} & \textbf{0.753} \\
%     \hline
%     \end{tabular}
%     \caption{Comparison between baseline and PFM over Flatvel B, CurveFault B, Style A and Style B.}
%     \label{tab:finetune_pretrain_model}
% \end{table}

% \begin{figure}
%     \centering
%     \includegraphics[width=\linewidth]{image/Baseline_vs_fullfinetune.png}
%     \caption{comparison between ground truth, baseline and full fine-tuning of the pretrained model on the four datasets : (a) FlatVel B, (b) CurveFault B, (c) Style A, and (d) Style B}
%     \label{fig:base_vs_FFT}
% \end{figure}
% \subsection{Comparison between the conventional and left-based finetuning method}

\subsection{Evaluation of the PEFT-based model }
% \dpk{Discuss here training and evaluation details of the pre-trained model and then evaluate its performance on our target tasks when approached through}\\
LoRA is a PEFT method commonly used for large models. It is gaining traction for its ability to adapt pre-trained models to specific tasks while minimizing resource usage. The training process using LoRA involves freezing the original weights of the PFM and only training a small number of additional parameters, which allows for efficient fine-tuning without the need for extensive computational resources \cite{benedek2024prilora}. The small number of additional parameters involves two trainable low-rank matrices to each layer, which are, LoRA A and LoRA B. LoRA A is trained to adapt features while B's initialization impacts stability and learning dynamics \cite{hayou2024impact}. The alpha $\alpha$ is the scaling factor between the determine the magnitude to which LoRA A and B affects the output. If we increase the alpha $\alpha$, the LoRA weight will dominate the output and if we decrease the LoRA weights then PFM dominate the output. To optimise the output, it is a common practice to fix alpha ($\alpha = 16$) \cite{lee2023platypus}.
The rank and alpha are two hyperparameters for LoRA, which are set to (rank, $r = 16$, alpha $\alpha=16$). 
The selection of rank and alpha have been performed by an exhaustive search over the range of values for $r$ = (4,8,16,32,64,128) and $\alpha$ = (4,8,16,32,64,128). 
%Table \ref{tab:rank_alpha} shows the performance of LoRA-PFM with different values of rank and alpha, however, it is a common practice to fix alpha ($\alpha = 16$) \cite{lee2023platypus}. 
Table \ref{tab:rank_alpha} shows the performance of LoRA-PFM with different values of rank and alpha. We conducted the same experiment using various combinations of rank and alpha values while keeping all other hyperparameters constant. This approach allowed us to identify the optimal settings that yielded the best performance across different tasks, highlighting the importance of hyperparameter tuning in achieving effective model adaptation.

%show the experimentation where you have decided the rank and alpha
\begin{table}[!h]
    \centering 
    \caption{\MakeUppercase{Selection of rank and alpha: Performance comparison for different rank values. The best-performing configuration, observed at rank = 16 and alpha = 16}}
    \begin{tabular}{|c|c|c|c|c|}
    \hline
    Rank & Alpha & MAE ($\downarrow$) & RMSE ($\downarrow$) & SSIM ($\uparrow$) \\
    \hline
    4 & 16 & 0.0674 & 0.131 & 0.883\\
    \hline
    8 & 16 & 0.0349 & 0.0911 & 0.936\\
    \hline
    16 & 16 & \textbf{0.030} & \textbf{0.080} & \textbf{0.952}\\
    \hline
    32 & 16 & 0.0327 & 0.0894 & 0.938\\
    \hline
    64 & 16 & 0.0441 & 0.099 & 0.9275 \\
    \hline
    128 & 16 & 0.0463 & 0.096 & 0.931 \\
    \hline
    \end{tabular}
    \label{tab:rank_alpha}
\end{table}

\begin{figure}[h]
    \centering
    \includegraphics[width=\linewidth]{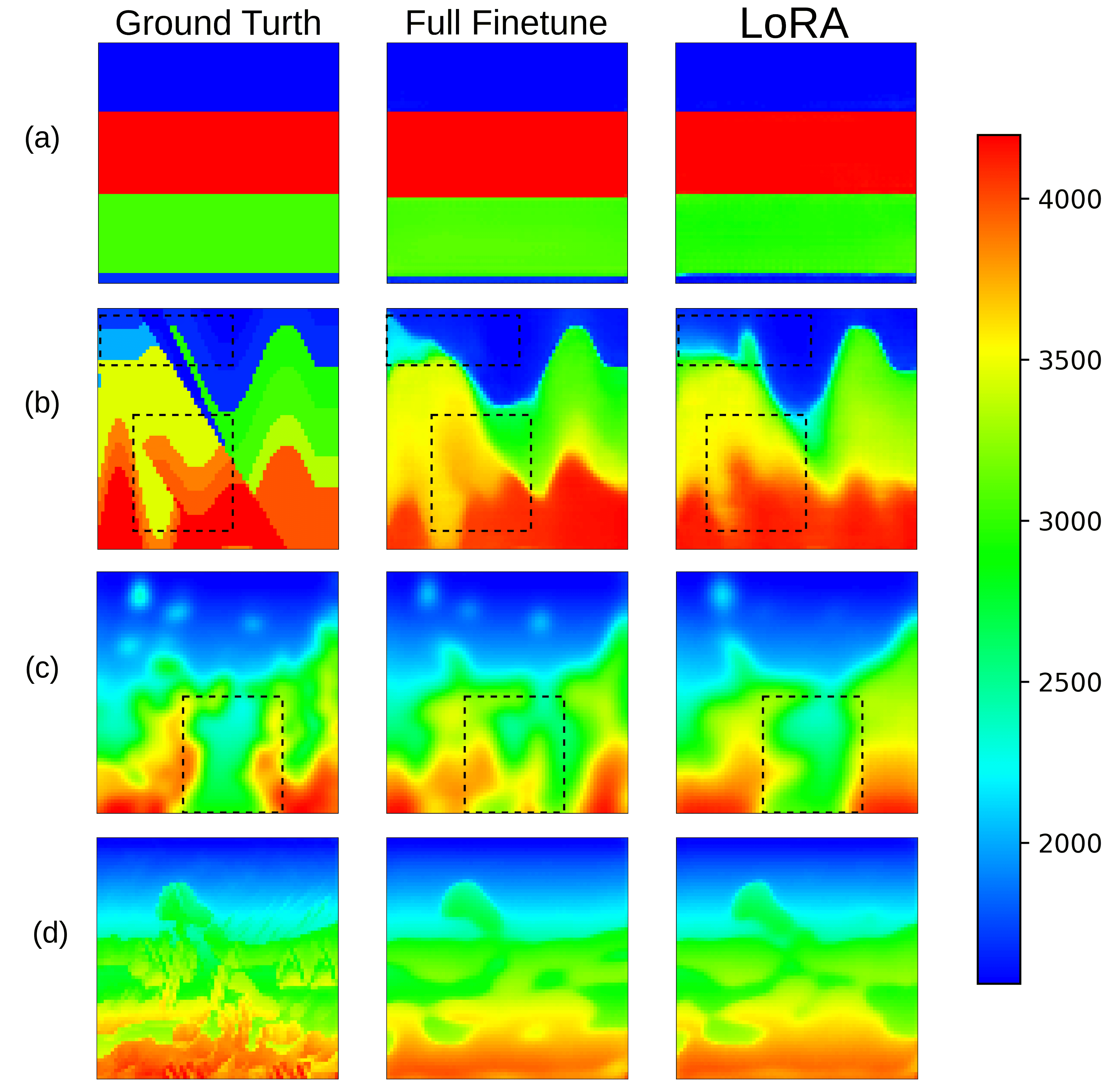}
    \caption{Comparison of full fine-tuning and LoRA-PFM over FlatVel B, CurveFault B, Style A, and Style B. Randomly selected two samples from the same dataset and (a), (b), (c), and (d) in the diagram represents  FlatVel B, CurveFault B, Style A and Style B respectively.}  \label{fig:full_data}
\end{figure}

\textbf{Evaluation: }We evaluate the performance of the model on both methods through qualitative and quantitative analysis. The quantitative analysis details are presented in Table \ref{tab:ODD_finetune}. It clearly shows that the FFT-PFM outperforms the baseline in all the datasets and is closely followed by LoRA-PFM. %However, the difference between FFT-PFM and LoRA-PFM is not statistically significant, indicating that both methods are effective in their own right. For FlatVel B and Style A, the difference in the improvement of the two methods is more pronounced than in the other two, suggesting that more complex datasets may benefit more from the LoRA-PFM of the pretrained model. However, full fine-tuning and LoRA-PFM are better than the baseline methods regarding MAE, RMSE, and SSIM. 
The qualitative analysis is shown in Fig. \ref{fig:full_data}; the recovered velocity maps obtained by FFT-PFM and LoRA-PFM are nearly the same. Fig. \ref{fig:full_data} shows two samples from the same dataset. 
In the case of FlatVel B, both FFT-PFM and LoRA-PFM were able to perfectly recover all the flat velocity layers, except the extra bottom layer, where LoRA-PFM estimated it accurately in comparison to FFT-PFM. Although Table \ref{tab:ODD_finetune} shows that FFT-PFM performs slightly better than LoRA-PFM from Fig. \ref{fig:full_data} we can observe that LoRA-PFM slightly performs better than FFT-PFM in deeper regions. 
% In the case of CurveFault B, we observe that the low-velocity layers and those faults that have a small change in velocity across them are not properly reflected in the predicted velocity map from both methods. Additionally, high-velocity deep layers are not accurately recovered by either method, but FFT-PFM is comparatively better at capturing some of the complex deeper variations than LoRA-PFM.
In the case of CurveFault B, we observe that the bounding boxes shown in Fig. \ref{fig:full_data}(b) are zones with significant improvement in velocity structures and values compared to FFT-PFM. LoRA-PFM was able to determine the fault's position and orientation accurately whereas FFT-PFM accurately determined the layer velocities.
For Style A, both FFT-PFM and LoRA-PFM can demarcate the layer boundary between lower velocity and higher velocity; however, LoRA-PFM fails to recover the small-scale bodies located at shallow depths. It accurately predicted the large-scale bodies at greater depths, whereas FFT-PFM predicts inaccurate velocity structures as shown by the bounding box in Fig. \ref{fig:full_data}(c).
Similarly, for Style B, the higher resolution velocity contrast is absent in the recovered velocity map for both methods, and the predicted velocity map shows a blurred representation of the true velocity map. 
Overall, we can conclude that the PFM performs better than the baseline with both the fine-tuning methods. Out of the two fine-tuning methods: FFT-PFM and LoRA-PFM, we observed that LoRA-PFM performs better than FFT-PFM in all the tested datasets. Quantitatively, FFT-PFM and LoRA-PFM scores slightly differ, whereas, in qualitative analysis, LoRA-PFM performs better than FFT-PFM.

% \begin{table*}[!h]
%     \centering 
%     \begin{tabular}{|c|c|c|c|c|c|}
%     \hline
%     Data & Method & param
%     & MAE ($\downarrow$) & RMSE ($\downarrow$) & SSIM ($\uparrow$) \\
%     \hline
%     \multirow{3}{*}{FlatVel\_B} & Baseline & 24.4 M & 0.035 & 0.087 & 0.946 \\
%     & FFT-PFM & 24.4 M & \textbf{0.030} & \textbf{0.080} & \textbf{0.952} \\
%     & LoRA-PFM & 1.1 M & 0.031 & 0.084 & 0.943 \\
%     \hline
%     \multirow{3}{*}{CurveFault\_B} & Baseline & 24.4 M & 0.164 & 0.247 & 0.616 \\
%     & FFT-PFM & 24.4 M & \textbf{0.139} & \textbf{0.217} & \textbf{0.654} \\
%     & LoRA-PFM & 1.1 M & 0.139 & 0.218 & 0.649 \\
%     \hline
%     \multirow{3}{*}{Style\_A} & Baseline & 24.4 M & 0.062 & 0.102 & 0.885 \\
%     & FFT-PFM & 24.4 M & \textbf{0.058} & \textbf{0.095} & \textbf{0.897} \\
%     & LoRA-PFM & 1.1 M & 0.061 & 0.099 & 0.885 \\
%     \hline
%     \multirow{3}{*}{Style\_B} & Baseline & 24.4 M & 0.068 & 0.161 & 0.631 \\
%     & FFT-PFM & 24.4 M & \textbf{0.056} & \textbf{0.091} & \textbf{0.753} \\
%     & LoRA-PFM & 1.1 M & 0.058 & 0.092 & 0.741 \\
%     \hline
%     \end{tabular}
%     \caption{Comparison of baseline, full fine-tune, and LoRA-PFM over the 4 different datasets, Flatvel B, CurveFault B, Style A and Style B. The trainable parameters along with the accuracy metrics demonstrate the efficiency of LoRA-PFM as an alternative fine-tuning method.}
%     \label{tab:ODD_finetune}
% \end{table*}
\begin{table}[!h]
    \caption{\MakeUppercase{Quantitative comparison of baseline, FFT-PFM, and LoRA-PFM over the 4 different datasets: FlatVel B, CurveFault B, Style A, and Style B. FVB, CFB, STA, STB, Params are the abbreviation for FlatVel B, CurveFault B, Style A, Style B and Parameters respectively.}}
    \centering
    \begin{tabularx}{\linewidth}{|X|X|X|X|X|X|}
    \hline
    Data & Method & Params & MAE ($\downarrow$) & RMSE ($\downarrow$) & SSIM ($\uparrow$) \\
    \hline
    \multirow{3}{*}{FVB} & Baseline & 24.4 M & 0.035 & 0.087 & 0.946 \\
    & FFT-PFM & 24.4 M & \textbf{0.030} & \textbf{0.080} & \textbf{0.952} \\
    & LoRA-PFM & 1.1 M & 0.031 & 0.084 & 0.943 \\
    \hline
    \multirow{3}{*}{CFB} & Baseline & 24.4 M & 0.164 & 0.247 & 0.616 \\
    & FFT-PFM & 24.4 M & \textbf{0.139} & \textbf{0.217} & \textbf{0.654} \\
    & LoRA-PFM & 1.1 M & 0.139 & 0.218 & 0.649 \\
    \hline
    \multirow{3}{*}{STA} & Baseline & 24.4 M & 0.062 & 0.102 & 0.885 \\
    & FFT-PFM & 24.4 M & \textbf{0.058} & \textbf{0.095} & \textbf{0.897} \\
    & LoRA-PFM & 1.1 M & 0.061 & 0.099 & 0.885 \\
    \hline
    \multirow{3}{*}{STB} & Baseline & 24.4 M & 0.068 & 0.161 & 0.631 \\
    & FFT-PFM & 24.4 M & \textbf{0.056} & \textbf{0.091} & \textbf{0.753} \\
    & LoRA-PFM & 1.1 M & 0.058 & 0.092 & 0.741 \\
    \hline
    \end{tabularx}
    \label{tab:ODD_finetune}
\end{table}

\subsection{Improved out-of-distribution (OOD) generalization}
In this section, we will evaluate the network's generalization ability by training it on one dataset and testing it on a dataset with a significant statistical distribution shift. This evaluation allows us to assess the robustness and adaptability of the model when faced with variations in data distribution. %, ultimately providing insights into its practical applicability in field situations.
We can mathematically define the OOD problem:  $\mathcal{X}$ as the input set and $\mathcal{Y}$ as the output space. The network is typically trained on a dataset $\mathcal{D}_{train}$ = ${(\text{x}_i,\text{y}_i)}^n_{i=1}$ sampled from the training distribution $\mathcal{P}_{train}(X,Y)$, where $\text{x}_i \in \mathcal{X}$ and $\text{y}_i \in \mathcal{Y}$. The test distribution $\mathcal{P}_{test}(X,Y)$ is different from the train distribution $\mathcal{P}_{train}(X,Y)$. The goal is to generalize over the unseen distribution, that is, minimize loss over the test dataset.

We have trained the network over FlatVel B and conducted inference over CurveFault B, Style A and Style B. Here CurveFault B, Style A and B are the OOD datasets that we use to evaluate the model's performance in scenarios that differ from the training conditions. Likewise, we trained the network with CurveFault B, Style A and Style B and used the remaining three datasets for testing. The network was additionally trained to employ three techniques: Baseline, FFT-PFM, and LoRA-PFM.

\textbf{Results: }Table \ref{tab:OOD_inference_100_per}, summarizes the experiments for all four datasets, which collectively shows that LoRA-PFM performs better than the Baseline and FFT-PFM. 
When trained on FlatVel B, LoRA-PFM demonstrates improved generalization across the other three datasets, leading to notable enhancements in MAE, RMSE, and SSIM. Similar patterns are seen when trained on CurveFault B and Style A. Conversely when trained on Style B, FFT-PFM slightly outperforms LoRA-PFM on FlatVel B and CurveFault B. However, LoRA-PFM achieves better MAE and RMSE results in Style A, while FFT-PFM has a slightly higher SSIM.
From Tables \ref{tab:ODD_finetune} and \ref{tab:OOD_inference_100_per}, we can observe that FFT-PFM is slightly better than LoRA-PFM when the network is finetuned on the same dataset it was trained on, known as the in-distribution(ID) domain. However, for OOD samples, LoRA-PFM performs better than FFT-PFM across all datasets. 
The comprehensive findings presented in Table \ref{tab:OOD_inference_100_per} indicate that LoRA-PFM surpasses both Baseline and FFT-PFM in OOD situations. These findings indicate that while FFT-PFM shows competitive performance for in-distribution scenarios, LoRA-PFM consistently excels in generalization and accuracy metrics across all the ODD datasets, highlighting its robustness as a preferred technique for diverse applications. 

% analysis of the visual results
The qualitative analysis presented in the Supplementary Figures, specifically Fig. S1(e), S2(e), S3(e), and S4(e), indicates that although the results are generally poor in the OOD, LoRA-PFM performs relatively better than FFT-PFM. When fine-tuned using FlatVel B, the predicted velocity maps for CurveFault B, Styles A and B, exhibit inaccuracies. However, LoRA-PFM reveals the relative bending of flat layers that is not observed in FFT-PFM. This difference is likely attributed to the influence of frozen weights in the PFM method utilized by LoRA-PFM. Fine-tuning with other datasets also shows similar results presented in the supplementary figures. 
%When fine-tuning over a more complex dataset like CurveFault B, the network shows better adaptability and improved generalization. In supplementary Fig. S2(e), we observe that LoRA-PFM was able to recover the flat layer velocity and undulating complex structures present in the shallow portion of the Style A and B dataset. Performance of LoRA-PFM gets significantly better when finetuned with the more complex datasets for example Style A and B. LoRA-PFM performance is better than FFT-PFM but the overall recovered velocity map is quite similar, suggesting that while both methods can capture the essential features of the datasets, LoRA-PFM offers an enhanced performance by accurately predicting the deeper layers and slightly better in recovering the structures of the velocity map.
The main takeaway from this section is the two fine-tuning methods FFT-PFM and LoRA-PFM show promising results. With FFT-PFM, the entire network is updated, while LoRA-PFM modifies only the delta weights (with a rank of 16 and alpha of 16). Both methods outperform the Baseline method in both ID and OOD scenarios. For ID, FFT-PFM is slightly ahead of LoRA-PFM in terms of MAE, RMSE, and SSIM. %but both methods yield nearly identical qualitative outcomes. 
However, LoRA-PFM demonstrates superior generalization to unseen data and outperforms FFT-PFM in OOD scenarios. This makes LoRA-PFM a more effective option for reducing generalization errors in OOD samples. Therefore, LoRA-PFM is considered a better fine-tuning method for minimizing these errors.
\begin{table}
    \centering
    \caption{\MakeUppercase{Quantitative comparison between the results of Baseline, FFT-PFM and LoRA-PFM on OOD dataset in terms of MAE, RMSE, and SSIM}}
    \begin{tabular}{|c|c|c|c|c|c|}
    %\begin{tabularx}{\linewidth}{|X|X|X|X|X|X|X|}
    \hline
    Train & Test & Method & MAE ($\downarrow$) & RMSE ($\downarrow$) & SSIM ($\uparrow$) \\
    \hline
    \multirow{9}{*}{FVB} & \multirow{3}{*}{CFB} & Baseline & 0.399 & 0.525 & 0.402 \\
    & & FFT-PFM & 0.321 & 0.434 & 0.420 \\
    & & LoRA-PFM & \textbf{0.271} & \textbf{0.374} & \textbf{0.471} \\
    \cline{2-6}
    & \multirow{3}{*}{STA} & Baseline & 0.258 & 0.346 & 0.585 \\
    & & FFT-PFM M & 0.202 & 0.272 & 0.624 \\
    & & LoRA-PFM & \textbf{0.164} & \textbf{0.223} & \textbf{0.678} \\
    \cline{2-6}
    & \multirow{3}{*}{STB} & Baseline & 0.224 & 0.283 & 0.519 \\
    & & FFT-PFM & 0.151 & 0.191 & 0.545 \\
    & & LoRA-PFM & \textbf{0.122} & \textbf{0.154} & \textbf{0.575} \\
    \hline

    \multirow{9}{*}{CFB} & \multirow{3}{*}{FVB} & Baseline & 0.425 & 0.579 & 0.472 \\
    & & FFT-PFM & 0.310 & 0.472 & 0.555 \\
    & & LoRA-PFM & \textbf{0.248} & \textbf{0.410} & \textbf{0.619} \\
    \cline{2-6}
     & \multirow{3}{*}{STA} & Baseline & 0.156 & 0.218 & 0.698 \\
    & & FFT-PFM & 0.137 & 0.191 & 0.720 \\
    & & LoRA-PFM & \textbf{0.119} & \textbf{0.166} & \textbf{0.743} \\
    \cline{2-6}
     & \multirow{3}{*}{STB} & Baseline & 0.118 & 0.154 & 0.589 \\
    & & FFT-PFM & 0.118 & 0.156 &  0.592 \\
    & & LoRA-PFM & \textbf{0.104} & \textbf{0.134} & \textbf{0.610} \\
    \hline

    \multirow{9}{*}{STA} & \multirow{3}{*}{FVB} & Baseline & 0.473 & 0.607 & 0.376 \\
    & & FFT-PFM & 0.358 & 0.478 & 0.440 \\
    & & LoRA-PFM & \textbf{0.264} & \textbf{0.386} & \textbf{0.511} \\
    \cline{2-6}
     & \multirow{3}{*}{CFB} & Baseline & 0.311 & 0.419 & 0.451 \\
    & & FFT-PFM & 0.280 & 0.379 & 0.457 \\
    & & LoRA-PFM & \textbf{0.256} & \textbf{0.349} & \textbf{0.474} \\
    \cline{2-6}
     & \multirow{3}{*}{STB} & Baseline & 0.092 & 0.130 & 0.670 \\
    & & FFT-PFM & \textbf{0.085} & 0.118 &  \textbf{0.680} \\
    & & LoRA-PFM & 0.088 & 0.118 & 0.667 \\
    \hline

    \multirow{9}{*}{STB} & \multirow{3}{*}{FVB} & Baseline & 0.543 & 0.679 & 0.350 \\
    & & FFT-PFM & 0.423 & \textbf{0.542} & \textbf{0.371} \\
    & & LoRA-PFM & \textbf{0.402} & 0.550 & 0.370 \\
    \cline{2-6}
     & \multirow{3}{*}{CFB} & Baseline & 0.283 & 0.368 & 0.432 \\
    & & FFT-PFM & \textbf{0.255} & \textbf{0.339} & \textbf{0.452} \\
    & & LoRA-PFM & 0.276 & 0.391 & 0.410 \\
    \cline{2-6}
     & \multirow{3}{*}{STA} & Baseline & 0.130 & 0.177 & 0.748 \\
    & & FFT-PFM & 0.117 & 0.162 &  \textbf{0.785} \\
    & & LoRA-PFM & \textbf{0.115} & \textbf{0.159} & 0.780 \\
    \hline
    \end{tabular}
    %\end{tabularx}
    \label{tab:OOD_inference_100_per}
\end{table}

\subsection{Improve performance in low data regime}
\label{sec: low data}
\begin{figure*}[!t]
    \centering
    \subfloat[Test: CurveFaultB]{
        \includegraphics[width=0.25\textwidth]{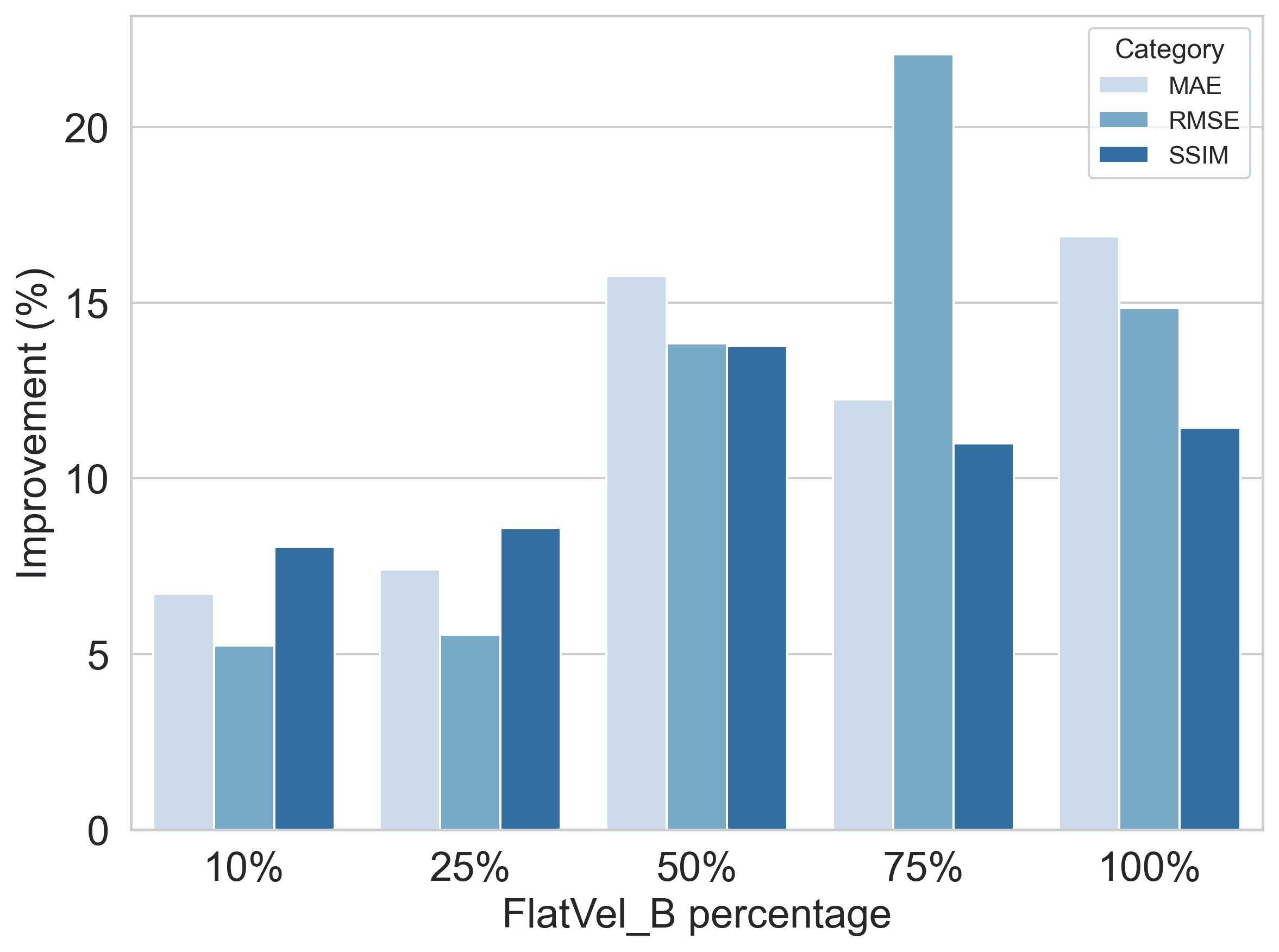}
        
    }
    \subfloat[Test: Style A]{
        \includegraphics[width=0.25\textwidth]{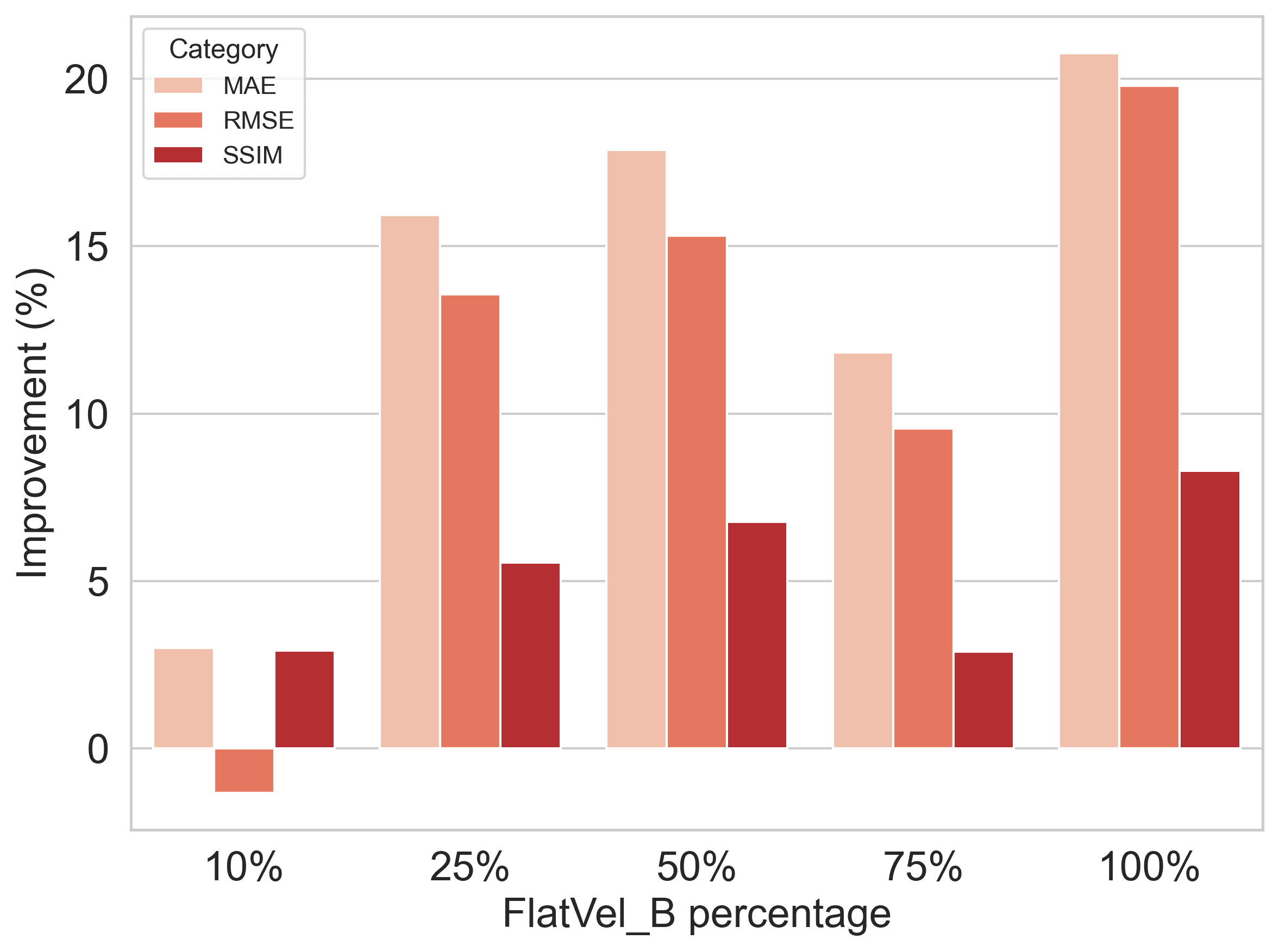}
        
    }
    \subfloat[Test: Style B]{
        \includegraphics[width=0.25\textwidth]{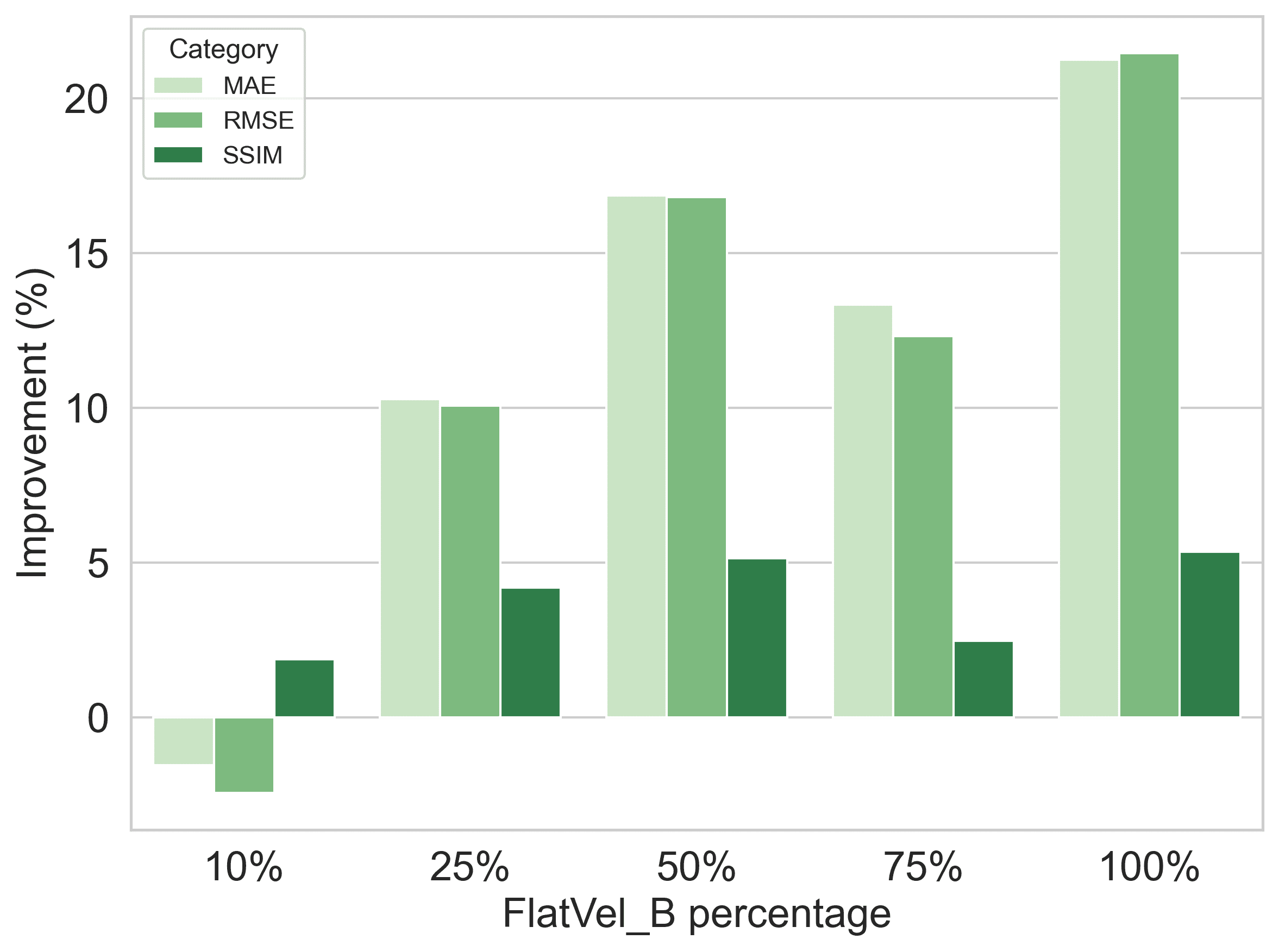}
        
    }
    \caption{Generalization improvement for fine-tuning with 10\%, 25\%, 50\%, 75\% and 100\% of training dataset and test with CurveFault B (blue), Style A (red), Style B (green). Evaluation was based on metric score, MAE, RMSE and SSIM. For MAE and RMSE, a lower value indicates better performance, while the reverse is true for SSIM.}
    \label{data_per_flatvelB}
\end{figure*}

\begin{figure*}
    \centering
    \includegraphics[width=\linewidth]{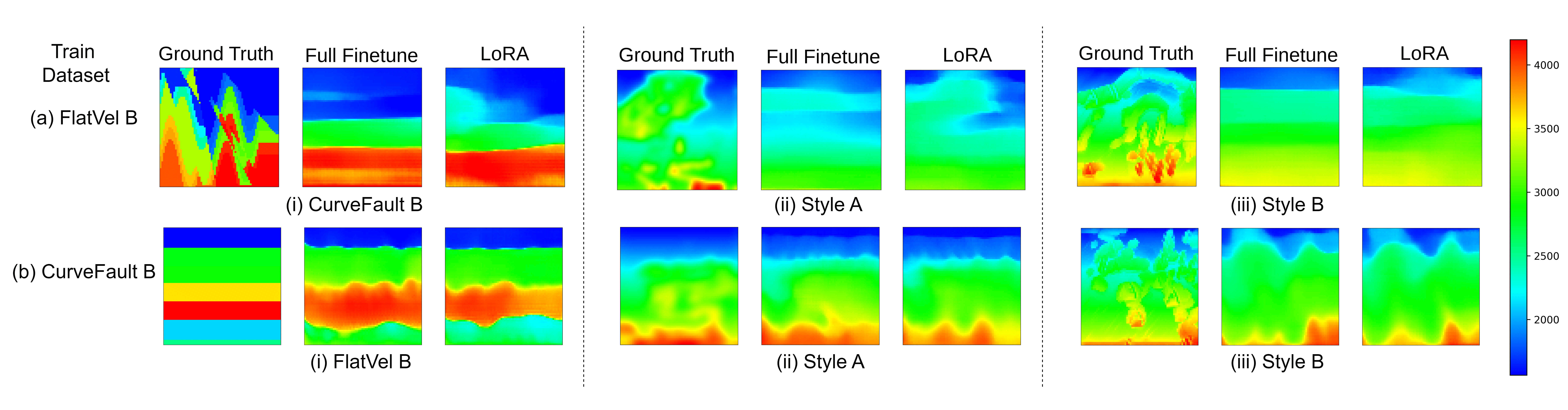}
    \caption{Predicted velocity map over different OOD datasets, trained with 10\% of the training dataset. We have evaluated the results for two different methods: LoRA-PFM and FFT-PFM. (a) and (b) shows the 10\% training dataset and  (i,ii, iii) represents the OOD test dataset. The predicted velocity maps are randomly selected from the test dataset.}
    \label{fig:ood_dat_per_train}
\end{figure*}

Generalization of data-driven methods struggles in low data regimes due to overfitting on small datasets. In this section, we demonstrate the benefit of LoRA-PFM in OOD generalization with low data availability. Here, we evaluate the same two methods with various dataset sizes (10\%, 25\%, 50\%, 75\% and 100\% of training dataset). In OpenFWI, each family contains a different number of samples, which are listed below:
\begin{enumerate}
    \item Vel family = 24000
    \item Fault family = 48000
    \item Style family = 60000
\end{enumerate}
%The evaluation metrics were assessed with varying sizes of the training dataset: 10\%, 25\%, 50\%, 75\% and 100\%. 
%This experiment was divided into two parts, the first part deals with the in-distribution and the second part deals with the OOD data. 
The PFM is fine-tuned with various percentages of the training dataset and tested on the OOD dataset. As mentioned in the previous section, the OOD dataset consists of three other datasets in addition to the one used for training. We examined the performance of both the proposed techniques (FFT-PFM and LoRA) in a low-data regime. Apart from the size of the dataset, all the other parameters remain the same throughout the experiment.\\
%All the hyperparameters and the size of the test dataset are the same throughout the experiment.\\
\textbf{Results:}\\
The results in Fig. \ref{data_per_flatvelB} demonstrate the improvement of PFM over FFT-PFM, when trained with FlatVel B and tested on CurveFault B, Style A and Style B. LoRA-PFM outperforms FFT-PFM significantly in terms of all the accuracy metrics. %indicating that the low-rank adaptation technique enhances the model's ability to generalize across different types and structures of the velocity map within the OOD test set. The bar in Fig. \ref{data_per_flatvelB} represents the performance improvement achieved by LoRA-PFM compared to FFT-PFM. 
We notice that as the dataset size increases, the performance improves in CurveFault B, Style A, and Style B. The test set is more complex than the training set, which means the network needs to learn more to adapt its weight for the test set. As the training dataset increases, the performance of the network improves with LoRA-PFM. FFT-PFM updates the weights of the entire network to learn the flat layers, but it predicts the flat layer incorrectly during inference with complex datasets (CurveFault B, Style A, and Style B). The frozen weight of PFM and the delta weights for Flatvel B together improve the performance of LoRA-PFM during inference with complex datasets. The results are shown in Fig. \ref{fig:ood_dat_per_train}(a), which depicts that LoRA-PFM exhibits more adaptability and generalization than the FFT-PFM model. However, all the results are inaccurate but LoRA-PFM was able to predict the shallow faults in the case of CurveFault B and shallow subsurface structures for Style A and Style B.

\begin{figure*}[!t]
    \centering
    \subfloat[Test: FlatVel B]{
        \includegraphics[width=0.25\textwidth]{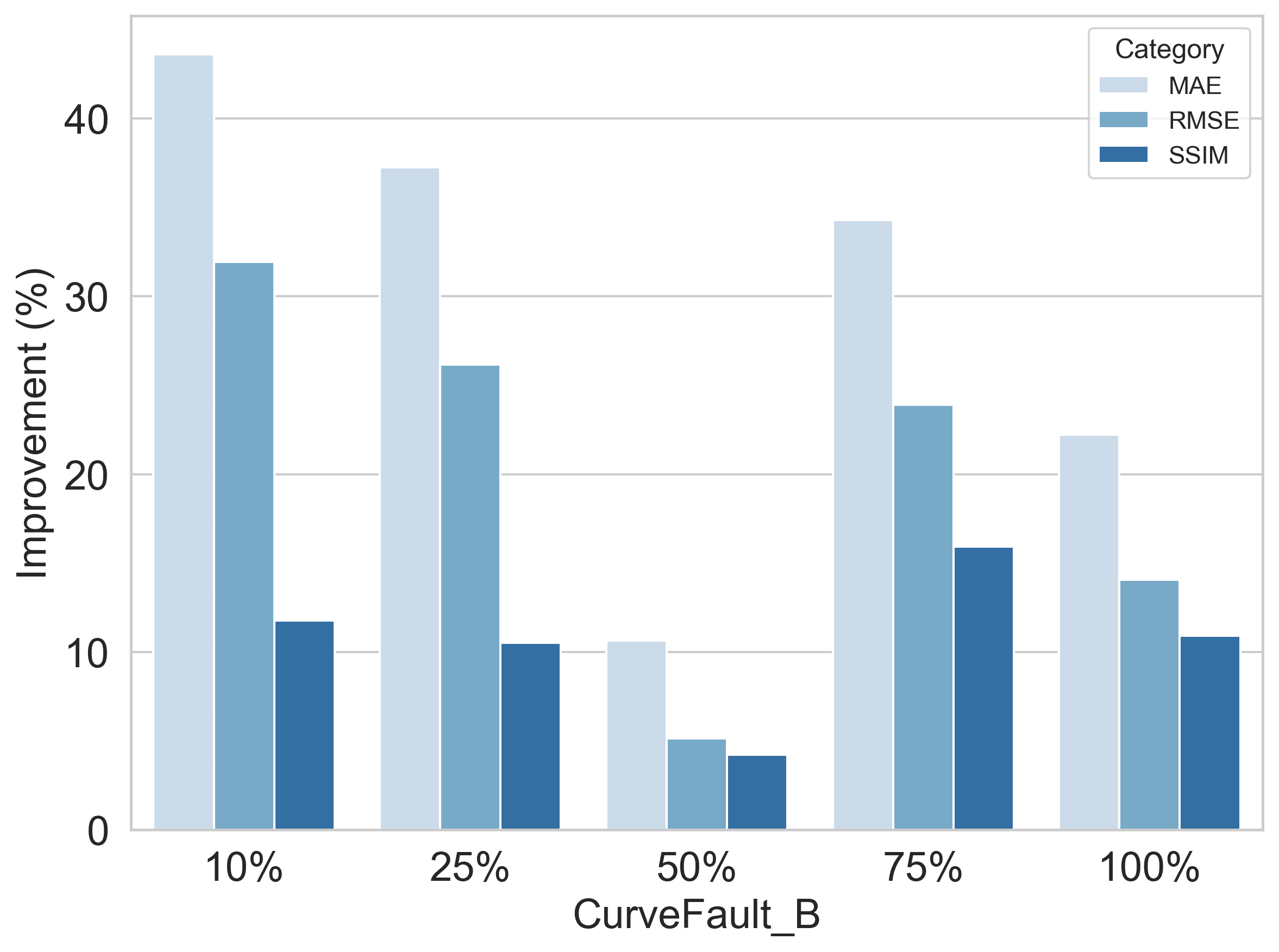}
        
    }
    \subfloat[Test: Style A]{
        \includegraphics[width=0.25\textwidth]{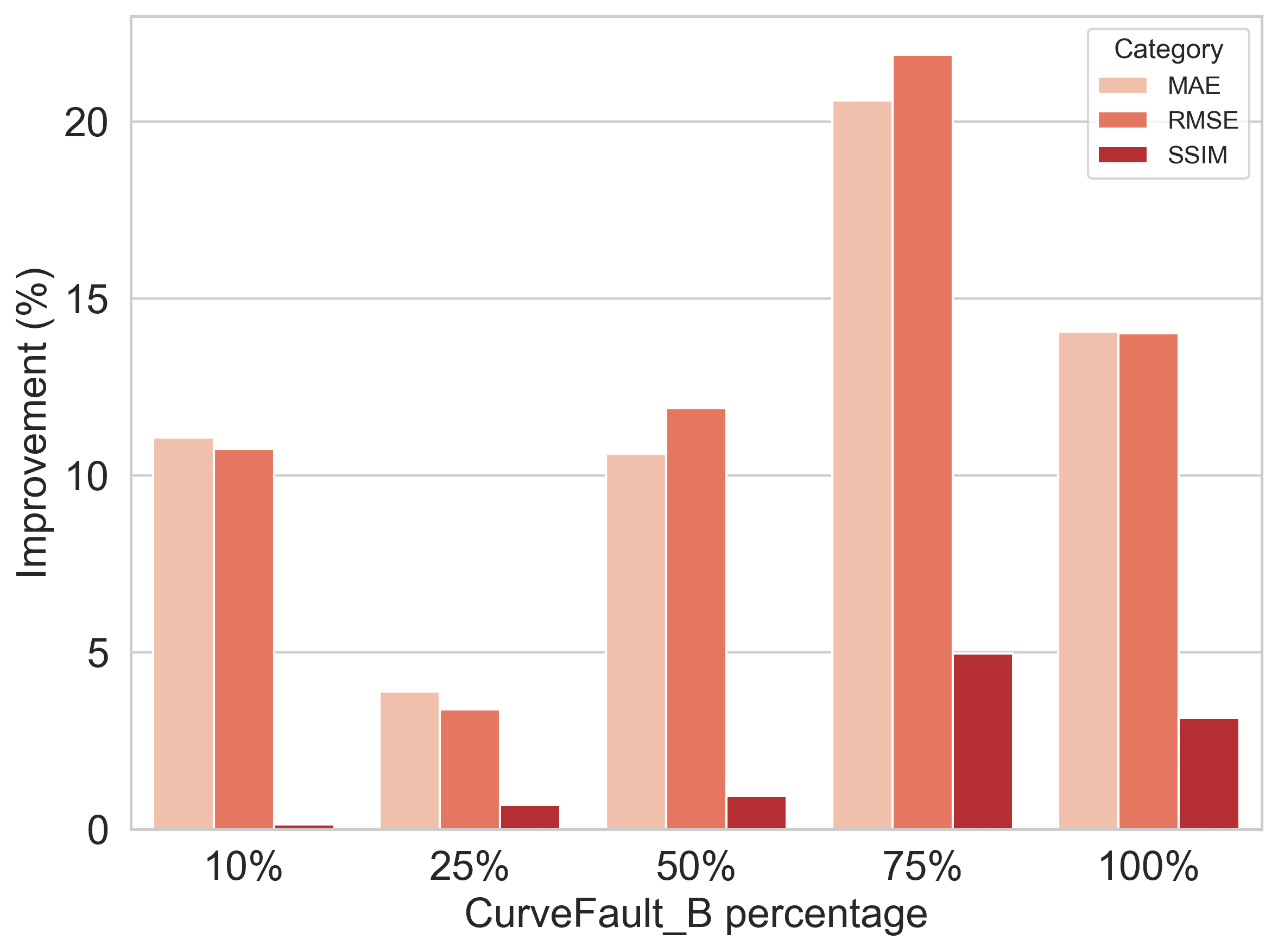}
        
    }
    \subfloat[Test: Style B]{
        \includegraphics[width=0.25\textwidth]{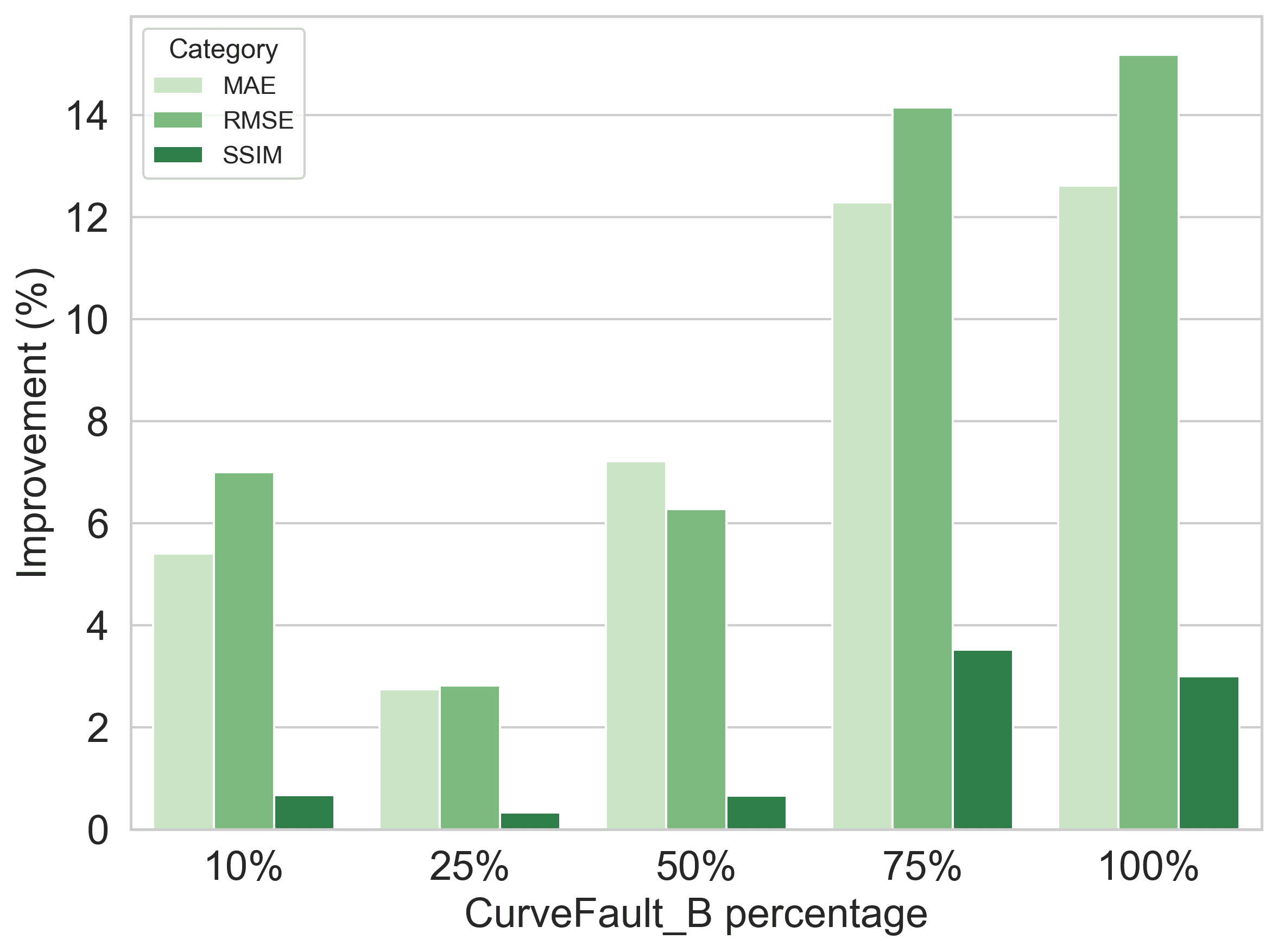}
        
    }
    \caption{Bar plot showing generalization improvement of LoRA-PFM over full finetuning. PFM was finetuned with various percentages of the training dataset (CurveFault B) and tested with FlatVel B (blue), and Style A (red).}
    \label{data_per_curvefaultB}
\end{figure*}

When PFM is fine-tuned with CurveFault B and tested with FlatVel B, Style A and Style B. Fig. \ref{data_per_curvefaultB} shows the improvement of LoRA-PFM in comparison to FFT-PFM with different sizes of the training dataset. During the inference over FlatVel B, we can observe that with increasing datasets the performance between FFT-PFM and LoRA-PFM converges. This pattern is observed because the training dataset contains complex structures and as FFT-PFM improves with increasing data, the difference between LoRA-PFM and FFT-PFM becomes small. For Style A and Style B, we observe the opposite trend in the improvement bar graph. As the dataset size increases, LoRA-PFM performs better than FFT-PFM. LoRA-PFM was able to learn useful features from the fine-tuning data, which resulted in high improvement with the increasing dataset. From Fig. \ref{fig:ood_dat_per_train}(b), we can observe that with 10\% of training data, LoRA-PFM was able to predict better than FFT-PFM in all the test datasets. Unlike the fine-tuning with FlatVel B, here both methods can capture the shallow complex structure accurately. However, inaccurate structure and predicted velocities are observed with depth. Both LoRA-PFM are relatively better than FFT-PFM in shallow parts as well as in deeper parts.

From this section, we can conclude that even in a low-data regime, LoRA-PFM was able to capture meaningful features from the small fine-tuning dataset. It effectively utilized the weights of both the PFM and LoRA modules to recover shallow structures during inference with out-of-distribution (OOD) samples. In contrast, FFT-PFM updates the entire network's weights, which tends to overfit the small fine-tuning dataset and performs poorly during inference. Based on all our evaluations, we conclude that LoRA-PFM is a better fine-tuning method compared to FFT-PFM for handling OOD samples. A detailed discussion of the fine-tuning results for Style A and Style B can be found in Section IV of the supplementary document.
% \begin{figure*}[!t]
%     \centering
%     \subfloat[Test: FlatVel B]{
%         \includegraphics[width=0.25\textwidth]{image/low_data/STA_FVB_barplot.png  }
%         \label{fig:SAFB}
%     }
%     \subfloat[Test: CurveFault B]{
%         \includegraphics[width=0.25\textwidth]{image/low_data/STA_CFB_barplot.png}
%         \label{fig:SACB}
%     }
%     \subfloat[Test: Style B]{
%         \includegraphics[width=0.25\textwidth]{image/low_data/STA_STB_barplot.png}
%         \label{fig:SASB}
%     }
%     \caption{Bar plot showing generalization improvement of LoRA-PFM over full finetuning. PFM was finetuned with various percentages of the training dataset (Style A) and tested with FlatVel B (blue), CurveFault B (red), and Style B (green).}
%     \label{data_per_StyleA}
% \end{figure*}
% \begin{figure*}[!t]
%     \centering
%     \subfloat[Test: FlatVel B]{
%         \includegraphics[width=0.25\textwidth]{image/low_data/STB_FVB_barplot.png  }
%         \label{fig:SBFB}
%     }
%     \subfloat[Test: CurveFault B]{
%         \includegraphics[width=0.25\textwidth]{image/low_data/STB_CFB_barplot.png}
%         \label{fig:SBCB}
%     }
%     \subfloat[Test: Style A]{
%         \includegraphics[width=0.25\textwidth]{image/low_data/STB_STB_barplot.png}
%         \label{fig:SBSA}
%     }
%     \caption{Bar plot showing generalization improvement of LoRA-PFM over full finetuning. PFM was finetuned with various percentages of the training dataset (Style B) and tested with FlatVel B (blue), CurveFault B (red), and Style A (green).}
%     \label{data_per_StyleB}
% \end{figure*}

\subsection{Efficient memory consumption of PEFT}
\label{sec: memory}
From the perspective of memory consumption, we observed that among three methods (Baseline, FFT-PFM, and LoRA), LoRA-PFM exhibited the lowest memory consumption for fine-tuning, as indicated in the Table \ref{tab:ODD_finetune}. Baseline and FFT-PFM utilize 24.4 million trainable parameters, whereas LoRA-PFM uses only 1.1 million parameters, which is a reduction of 4.5\%. Despite using a significantly lower number of parameters, we observe the same performance between FFT-PFM and LoRA-PFM in the case of in-distribution samples, whereas in the case of OOD, LoRA-PFM outperforms FFT-PFM. It indicates that LoRA-PFM is an efficient fine-tuning technique in terms of memory consumption. Further, as the model size increases, the efficiency of the LoRA-PFM becomes more apparent.

\section{Conclusion}
We demonstrated how building a pretrained foundational model and then fine-tuning it with PEFT can deliver significantly improved performance on FWI tasks compared to the conventional approach. Moreover, our fine-tuning memory is efficient, which in turn adapts to several different
This paper addresses the limitations of task-specific models in seismic full waveform inversion (FWI) by introducing a task-agnostic foundational model combined with a parameter-efficient fine-tuning (PEFT) method. The foundational model (PFM), pretrained on a diverse dataset from OpenFWI, captures generalized features across various geological scenarios. PFM captures essential features from its large-scale and cross-domain training dataset.  We trained the InversionNet with 6 datasets collected from OpenFWI and fine-tuned it over 4 datasets. Through our experiment, we have demonstrated that PFM achieves high accuracy over the complex velocity maps in comparison to the baseline model.

The study also demonstrates that full fine-tuning of the PFM outperforms task-specific models, and PEFT methods, particularly Low-Rank Adaptation (LoRA), achieve comparable results with significantly reduced computational and memory requirements. We found that LoRA-PFM takes advantage of both worlds, as it harnesses the power of PFM along with the flexibility to adapt them efficiently to new tasks. These qualities help LoRA-PFM to excel in OOD tasks and low-data regimes, enhancing generalization and performance. However, in the case of ID LoRA-PFM scores are not as better as FFT-PFM. However, LoRA-PFM can be very beneficial in the case of large models; other PEFT methods can be more beneficial depending on the problem at hand. With the increasing growth of efficient fine-tuning algorithms, this study will facilitate the seamless integration and assessment of novel PEFT techniques across various seismic applications in the future. 

Our study relies entirely on OpenFWI, which offers convenience but has inherent limitations. We observe a noticeable gap between synthetic data and field data. As a result, our experiments are confined to simulations. Addressing this gap remains an ongoing challenge for the entire FWI community, requiring efforts to make more public field data available or enhance the realism of simulations. Recent developments like Fourier-DeepONet \cite{zhu2023fourier} made some effort to increase the generalization ability of the data-driven DL-based FWI. However, challenges persist, encouraging the search for a more generalized method for FWI. This paper demonstrates the potential of combining PFMs with PEFT techniques to enhance generalization, improve performance in ID and OOD, along with low-data regimes in OOD, and reduce computational and memory costs in DL-FWI. While our study specifically focuses on FWI, the principles and methods presented here are not limited to this task alone. The demonstrated effectiveness of PEFT can be extended to other geophysical challenges, paving the way for broader applications in seismic interpretation, reservoir characterization, and beyond.
%Since  there is a significant gap between the simulated data and the field data. The data-driven DL-based FWI struggles with field datasets that have significant variation from the training dataset. 
%The paper highlights the potential of combining foundational models with PEFT to create efficient, adaptable models for diverse geophysical tasks. This makes the proposed method especially significant for field application. 

\section*{Acknowledgments}
The authors would like to thank Nyun AI for providing the computing infrastructure needed for the experiments conducted in this paper.

\section{Data and Codes Availability}
OpenFWI data set can be downloaded from the website (https://openfwi-lanl.github.io/). Pretraining and PEFT codes
are released and can be downloaded from the Website (https://github.com/Kaustav546/FWI-PEFT.git).

\bibliographystyle{IEEEtran}
\bibliography{sample}
\newpage

% \documentclass[journal]{IEEEtran}
% \usepackage{amsmath,amsfonts}
% \usepackage{algorithmic}
% \usepackage{algorithm}
% \usepackage{amssymb}
% \usepackage{array}
% \usepackage[caption=false,font=normalsize,labelfont=sf,textfont=sf]{subfig}
% \usepackage{textcomp}
% \usepackage{tabularx}
% \usepackage{titling}
% \usepackage{import}

% \usepackage{stfloats}
% \usepackage{url}
% \usepackage{multirow}
% \usepackage{verbatim}
% \usepackage{graphicx}
% \usepackage{cite}
% \usepackage{lineno}
% \usepackage{amssymb}
% \usepackage{caption}
% \usepackage{subcaption}
% \usepackage{longtable}
% \usepackage[normalem]{ulem}
% \usepackage{subfiles}
% \usepackage{xcolor}

% \documentclass[lettersize,journal]{IEEEtran}
% \centering
\title{(Supplementary document)}
% \date{} % Specify a custom date



\maketitle
\date{}
\author{}
\section{Implementation Details}
\label{ID}
\textbf{Loss and Evaluation metrics:}
Pretrained InversionNet utilizes the L1 norm as its loss function, $\textsc{L1-norm} = |y_p - y_o|/N$, where $N$ is the total number of data points in the velocity maps. It measures the discrepancy between the predicted velocity and the observed velocity map. We have used three evaluation metrics, MAE, RMSE and SSIM. Both MAE (same as the L1  norm) and $\textsc{RMSE} = \sqrt{|y_p - y_o|^2/N}$ capture the numerical difference between the predicted and observed velocity maps. $ \text{SSIM}(x, y) = \frac{(2 \mu_x \mu_y + C_1)(2 \sigma_{xy} + C_2)}{(\mu_x^2 + \mu_y^2 + C_1)(\sigma_x^2 + \sigma_y^2 + C_2)}
$ is used to measure the similarity between the two velocity maps. While measuring the MAE and RMSE, the velocity map is normalized between [-1, 1] but in the case of SSIM, we rescale it to [0,1].\\
\textbf{Hyperparameters:}
The convergence was achieved after 90 epochs in pretraining. The entire process takes approximately 190 hours, with the distributed training of over 8 Nvidia L4 GPUs, each with 24 GB of memory. The total size of the combined dataset used for pretraining is 317 GB. To optimize the network we have used the AdamW optimizer with beta = (0.9, 0.999) and WarmupMultiStepLR scheduler with warmup factor set to $1e^{-5}$.  The initial learning rate is set to $8\times10^{-4}$, and the models undergo training for 120 epochs. During the first five warm-up epochs, we gradually raise the learning rate from $1\times10^{-4}$, and we reduce the learning rate by a factor of 10 at epoch 90 and epoch 100, respectively. We set the batch size to 128 and used natural logarithmic transformation to balance the intensity of the seismic data, normalizing the seismic data and levels to [-1, 1].
\section{Performance evaluation of the foundational model}
\label{PFM_vs_Baseline}
\begin{figure}
    \centering
    \includegraphics[width=\linewidth]{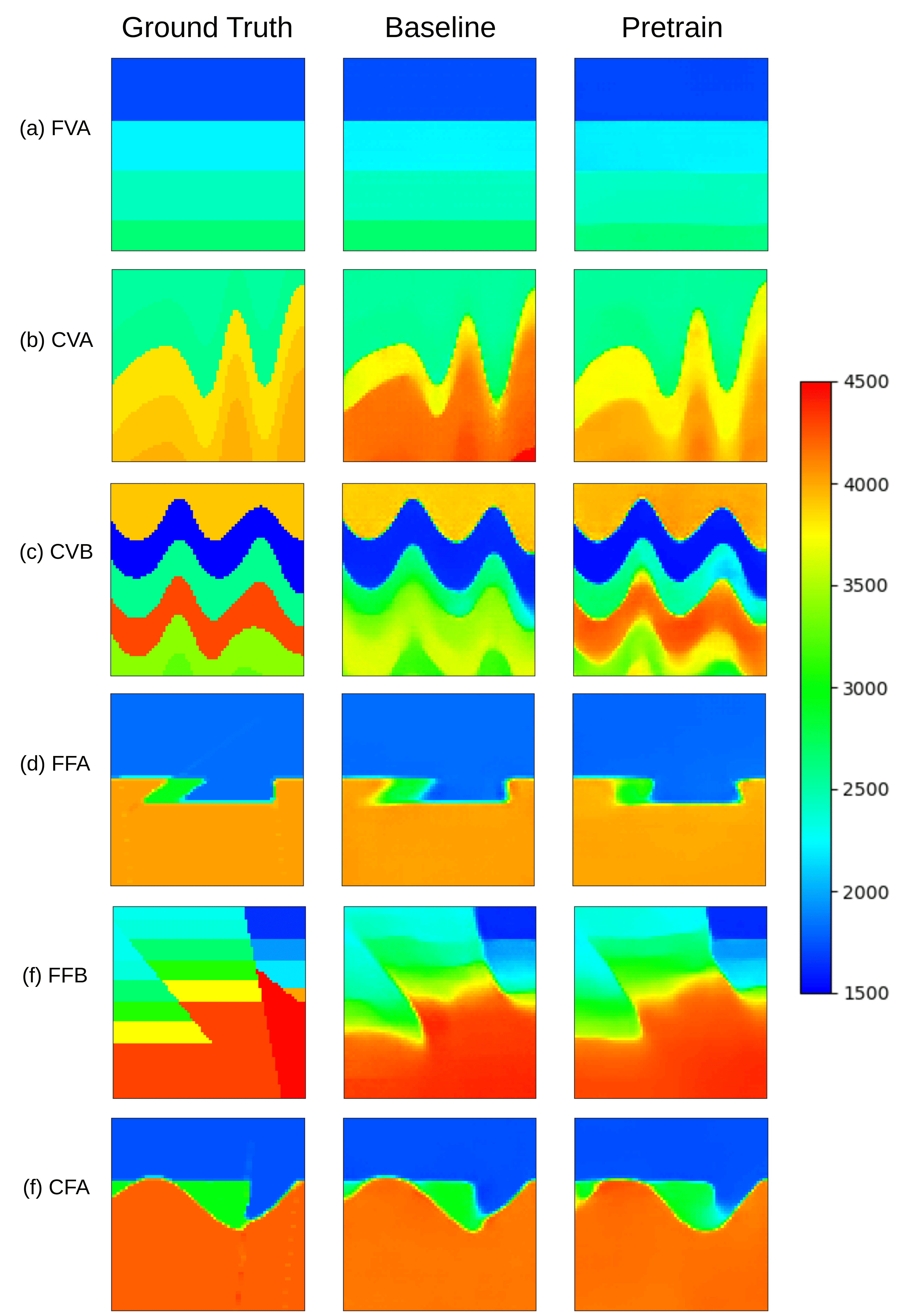}
    \caption{Comparison of the baseline and the PFM with the ground truth on the validation set. Here we randomly selected samples from the validation set.}
    \label{fig:pretrain_base}
\end{figure}

The Fig. \ref{fig:pretrain_base} shows the result of PFM and Baseline. The figure illustrates how the models benefit from pretraining in terms of feature extraction and accurately recovering the velocity map. We observe that pretraining significantly improves over more challenging datasets like CurveVel A, CurveVel B, and FlatFault B in comparison to simple datasets such as FlatVel A, FlatFault A, and CurveFault A show minimal improvements. This suggests that pretraining results in enhanced feature extraction, which allows the model to capture intricate patterns within the data. This ultimately leads to greater robustness when facing complex or OOD samples. That’s why we observe better performance in complex datasets in comparison to simple datasets. For a simple dataset, the baseline model overfits and results in more accurate predictions concerning PFM. For CurveVel A, the baseline model predicted much higher velocities in the deeper layers, whereas the pre-trained model (PFM) successfully recovered the velocities in those deeper layers.  Similarly, for CurveVel B, the baseline model fails to predict the correct velocities in the high-velocity layer situated in the deeper region. In the other data types, we observed similar results for both methods, although the visual improvement was marginal. Overall, we can conclude that pretraining enhances the network's performance on complex datasets and improves its ability to generalize across various geological scenarios. This leads to more reliable interpretations and better predictive performance in subsurface modelling.

\section{Finetuning of the foundational model}
\begin{figure}
    \centering
    \includegraphics[width=\linewidth]{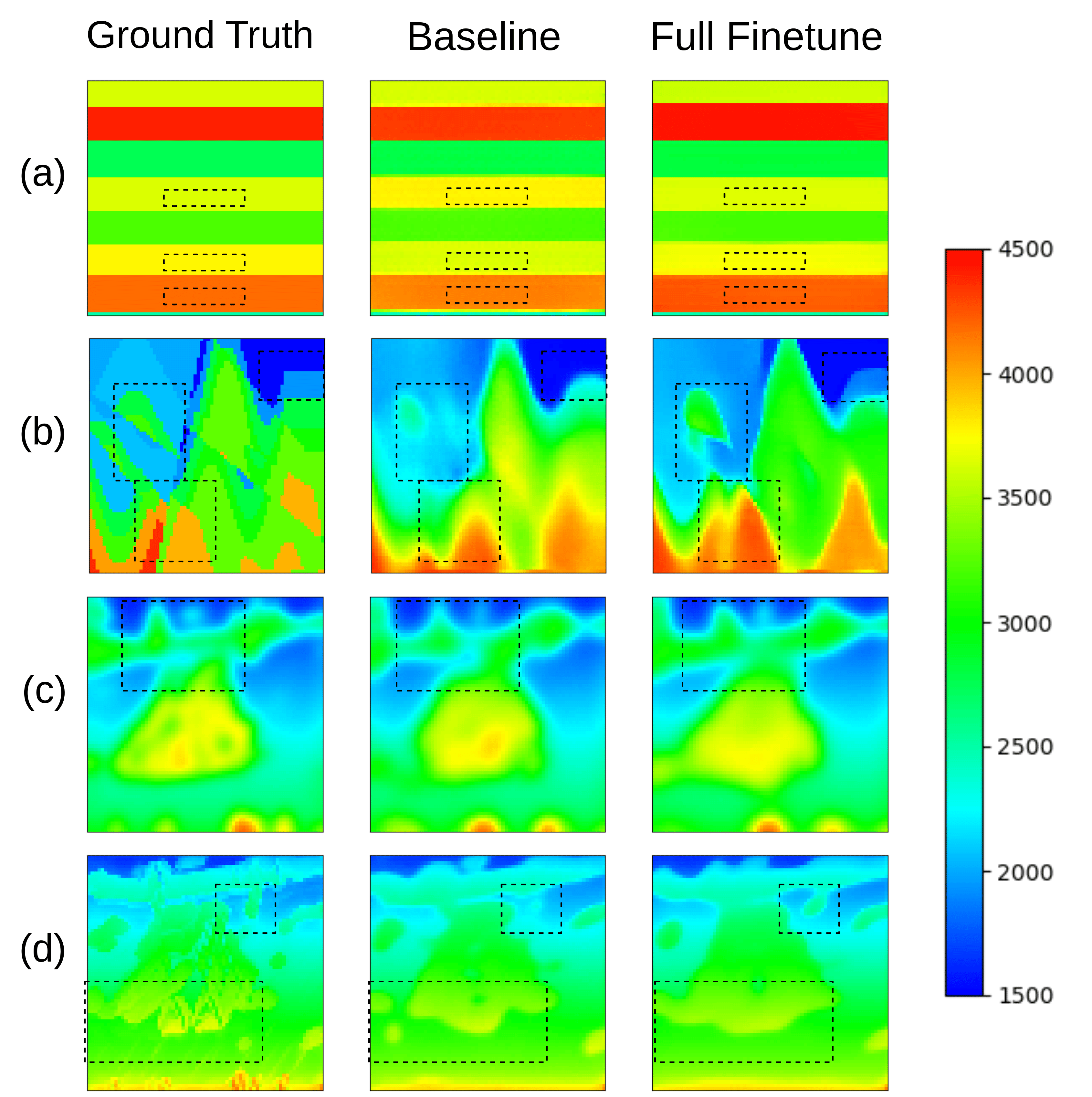}
    \caption{comparison between ground truth, baseline and full fine-tuning of the pretrained model on the four datasets : (a) FlatVel B, (b) CurveFault B, (c) Style A, and (d) Style B}
    \label{fig:base_vs_FFT}
\end{figure}

Fig. \ref{fig:base_vs_FFT} illustrates the true and predicted velocity distribution derived from both the FFT-PFM and Baseline methods. Our statistical analysis, presented in Table \ref{tab:finetune_pretrain_model}, indicates that fine-tuning the PFM results in substantial enhancements across key performance metrics, namely MAE, RMSE, and SSIM scores. As shown in Fig. \ref{fig:base_vs_FFT}(a), it is evident that the FFT-PFM method produces a more accurate velocity map, whereas the baseline approach fails to accurately predict the velocity values in the specifically marked layers highlighted in the same figure.
A closer examination of Fig. \ref{fig:base_vs_FFT}(b) reveals that the faults situated in the emphasized regions are effectively recovered using the FFT-PFM method. In contrast, the baseline technique significantly struggles to maintain the structural integrity necessary for an accurate representation. Notably, a significant performance enhancement is particularly evident in CurveFault B. Although the recovered velocity maps for Style A and Style B still display minor discrepancies when compared to the ground truth, these variations are identifiable in the designated areas of the velocity map.

\begin{table}[!h]
    \centering 
    \caption{Comparison between baseline and PFM over Flatvel B, CurveFault B, Style A and Style B.}
    \begin{tabular}{|c|c|c|c|c|c|}
    \hline
    Data & Method
    & MAE ($\downarrow$) & RMSE ($\downarrow$) & SSIM ($\uparrow$) \\
    \hline
    \multirow{2}{6em}{FlatVel\_B} & Baseline & 0.035 & 0.087 & 0.946 \\
    & FFT-PFM & \textbf{0.030} & \textbf{0.080} & \textbf{0.952} \\
    \hline
    \multirow{2}{6em}{CurveFault\_B} & Baseline & 0.164 & 0.247 & 0.616 \\
    & FFT-PFM & \textbf{0.139} & \textbf{0.217} & \textbf{0.654} \\
    \hline
    \multirow{2}{6em}{Style\_A} & Baseline & 0.062 & 0.102 & 0.885 \\
    & FFT-PFM & \textbf{0.058} & \textbf{0.095} & \textbf{0.897} \\
    \hline
    \multirow{2}{6em}{Style\_B} & Baseline & 0.068 & 0.091 & 0.631 \\
    & FFT-PFM & \textbf{0.056} & \textbf{0.082} & \textbf{0.753} \\
    \hline
    \end{tabular}

    \label{tab:finetune_pretrain_model}
\end{table}

\section{Low data regime}
\begin{figure*}[!t]
    \centering
    \subfloat[Test: FlatVel B]{
        \includegraphics[width=0.25\textwidth]{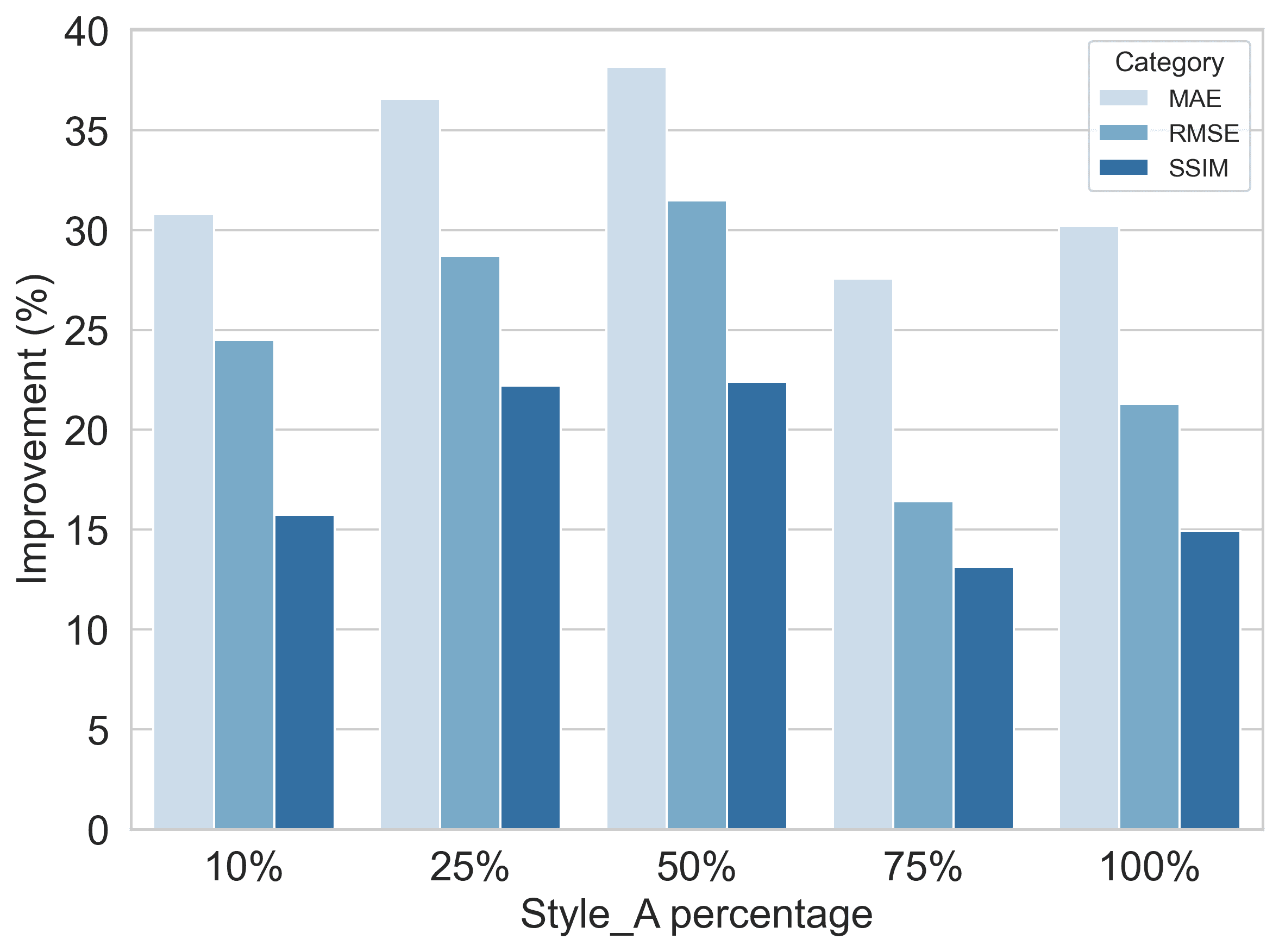  }
        \label{fig:SAFB}
    }
    \subfloat[Test: CurveFault B]{
        \includegraphics[width=0.25\textwidth]{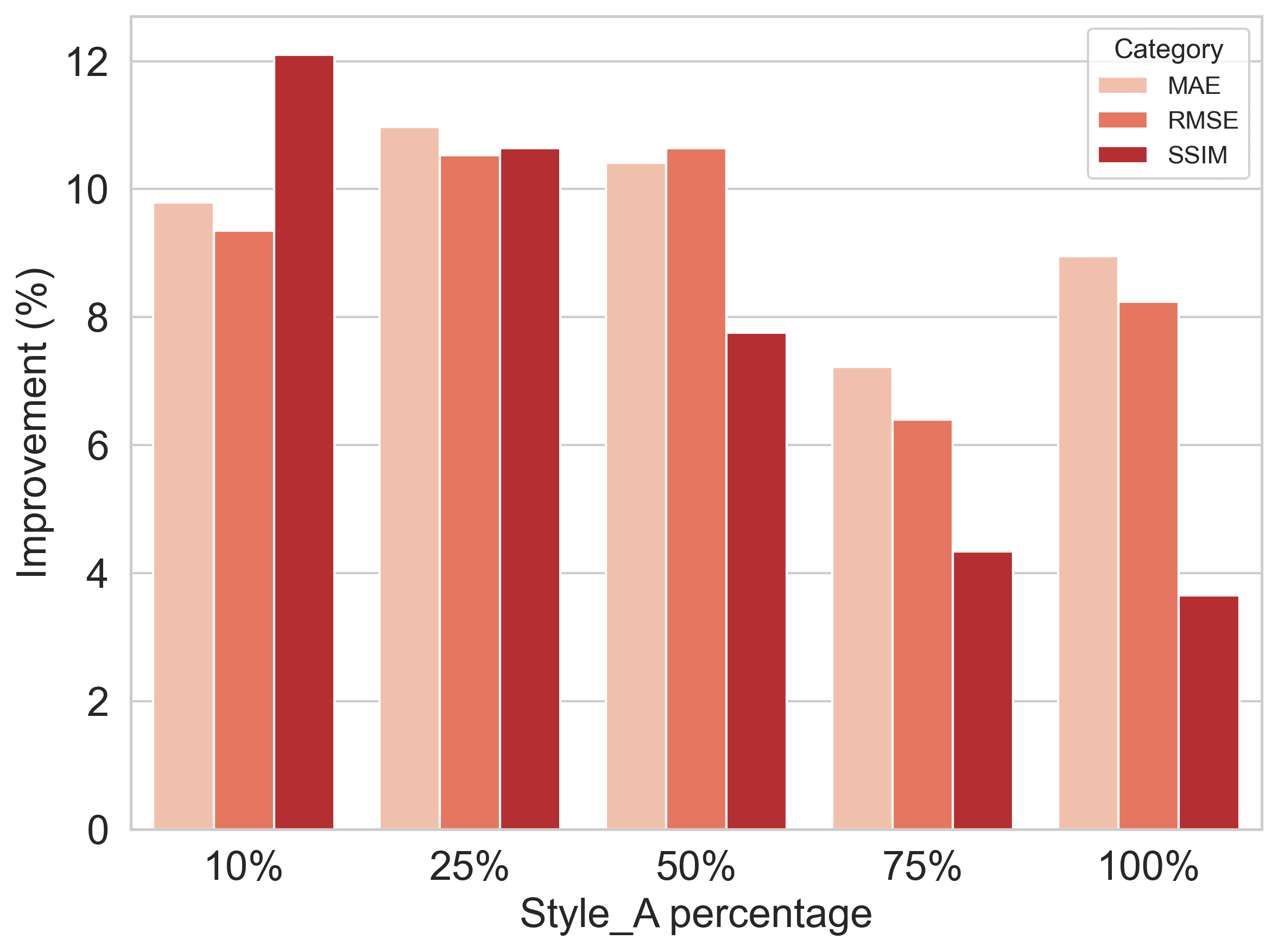}
        \label{fig:SACB}
    }
    \subfloat[Test: Style B]{
        \includegraphics[width=0.25\textwidth]{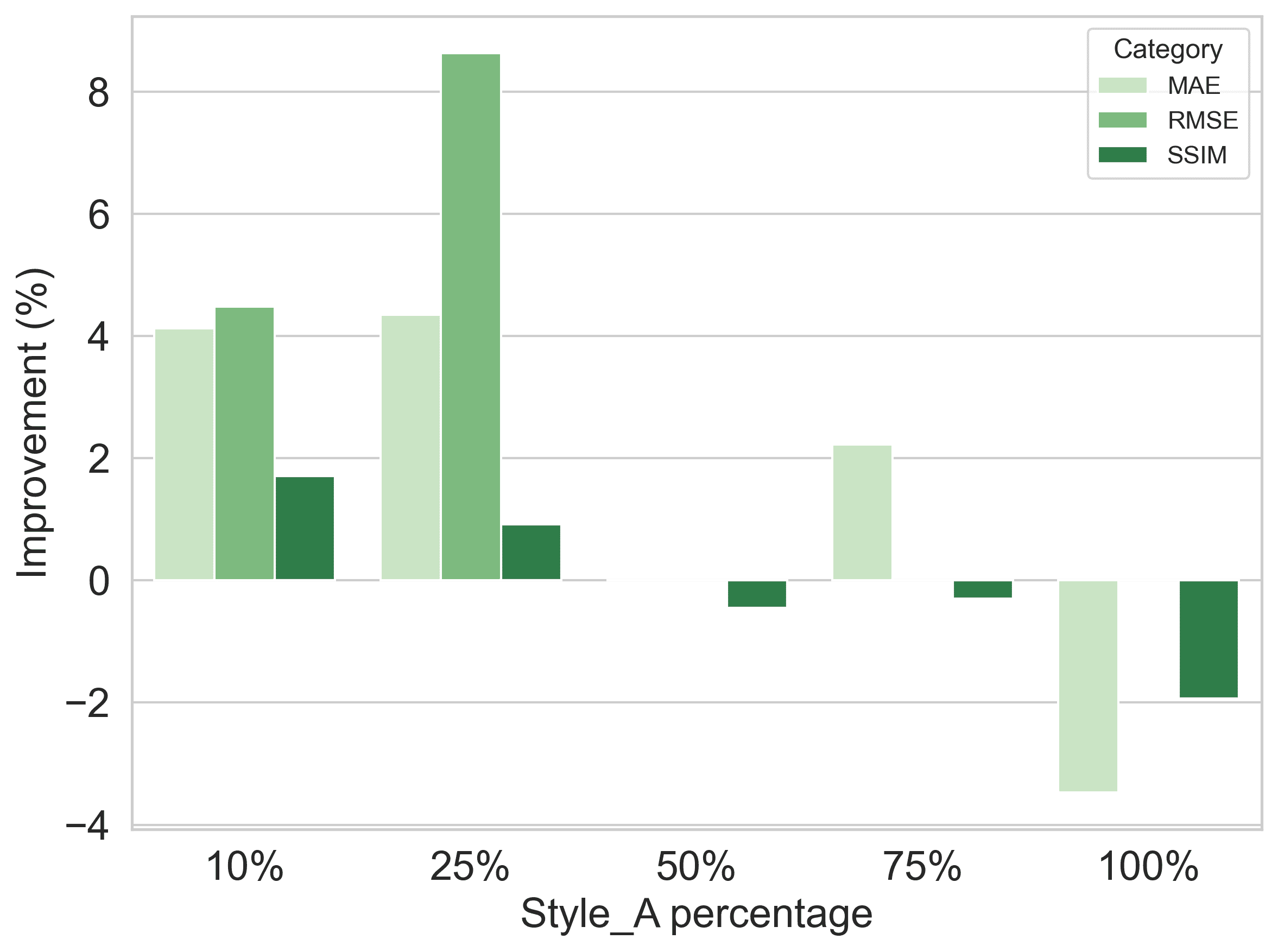}
        \label{fig:SASB}
    }
    \caption{Bar plot showing generalization improvement of LoRA-PFM over full finetuning. PFM was finetuned with various percentages of the training dataset (Style A) and tested with FlatVel B (blue), CurveFault B (red), and Style B (green).}
    \label{data_per_StyleA}
\end{figure*}

LoRA-PFM outperforms FFT-PFM in all metric scores across the entire OOD test set, indicating better generalization.
Fig. \ref{data_per_StyleA} shows the improved performance of LoRA-PFM in comparison with FFT-PFM when the PFM is fine-tuned with Style A and tested with FlatVel B, CurveFault B, and Style B. During inference with FlatVel B, LoRA-PFM performs better than FFT-PFM with every varying percentage of dataset size as shown in Fig. \ref{data_per_StyleA}(a). For CurveFault B, the improvement of LoRA-PFM and FFT-PFM converge as the dataset increases from 50\%. Compared to the high improvement percentage in FlatVel B, CurveFault B has a small improvement percentage. This is due to the presence of complex structures in the test dataset and the presence of dissimilar features in the training dataset and thus inaccurate predictions are made by both methods as shown in Fig. \ref{fig:ood_dat_per_train}(c). The similarity between Style A and B yields accurate prediction by both LoRA-PFM and FFT-PFM as shown in Fig. \ref{fig:ood_dat_per_train}(c). Due to the good performance of FFT-PFM with Style B, the improvement of LoRA-PFM with respect to FFT-PFM is comparatively smaller and with increasing dataset size we observe a crossover after 50\% (fig. \ref{data_per_StyleA}(c)).
%The difference can be visualized in Fig. \ref{fig:ood_dat_per_train}(c), although the recovered velocity maps are inaccurate LoRA-PFM was able to reduce the error in comparison to FFT-PFM.

\begin{figure*}[!t]
    \centering
    \subfloat[Test: FlatVel B]{
        \includegraphics[width=0.25\textwidth]{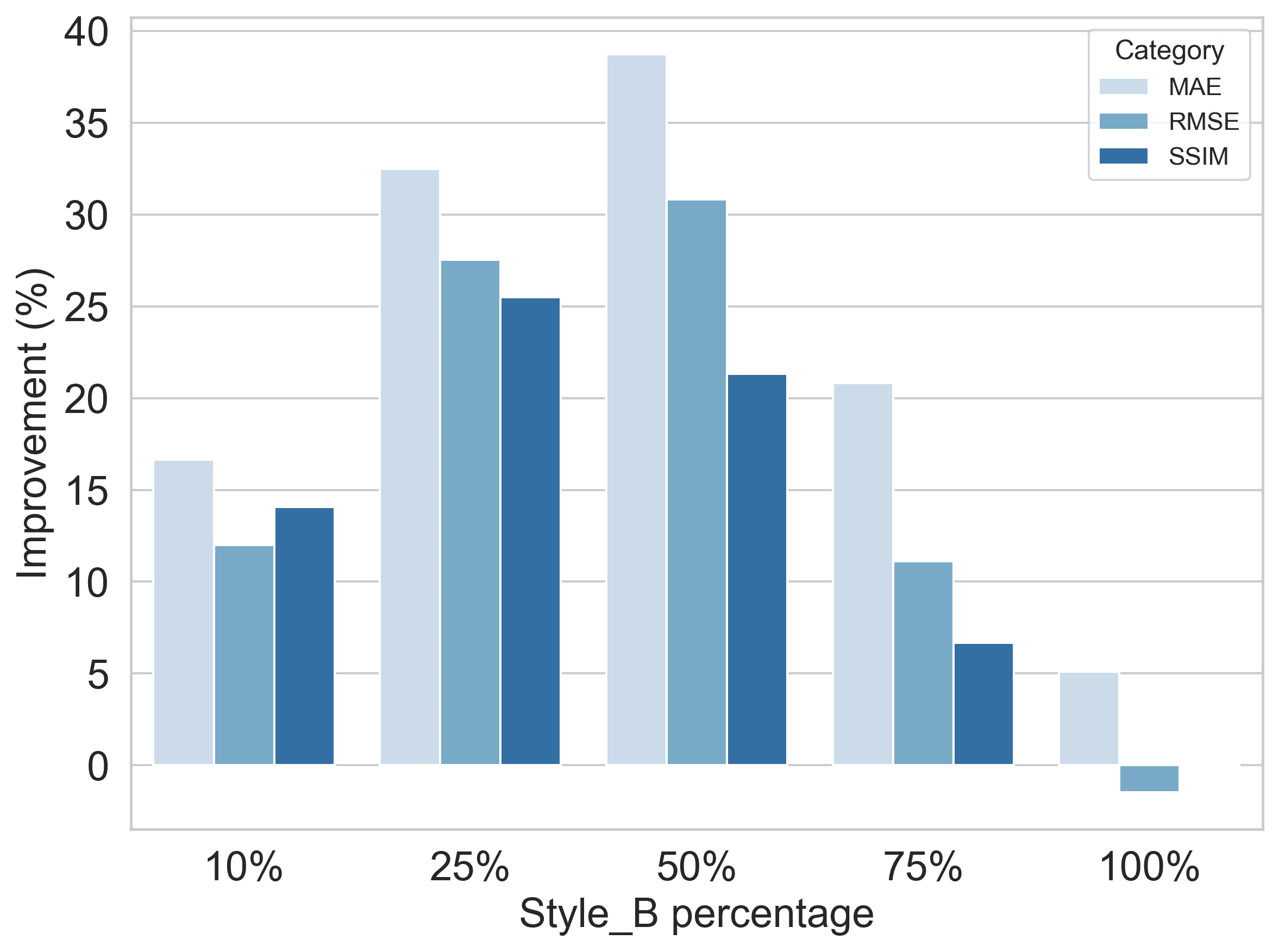  }
        \label{fig:SBFB}
    }
    \subfloat[Test: CurveFault B]{
        \includegraphics[width=0.25\textwidth]{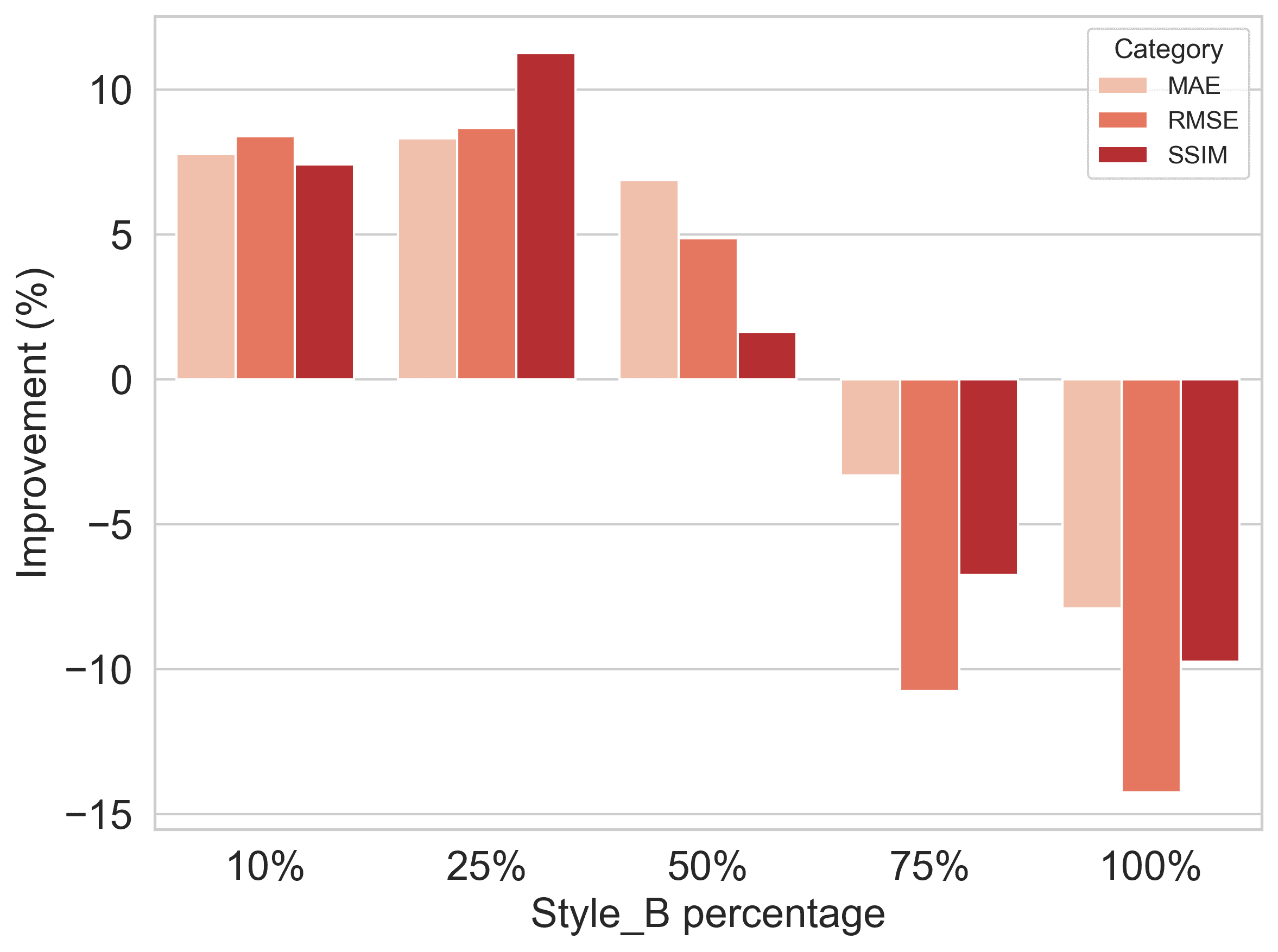}
        \label{fig:SBCB}
    }
    \subfloat[Test: Style A]{
        \includegraphics[width=0.25\textwidth]{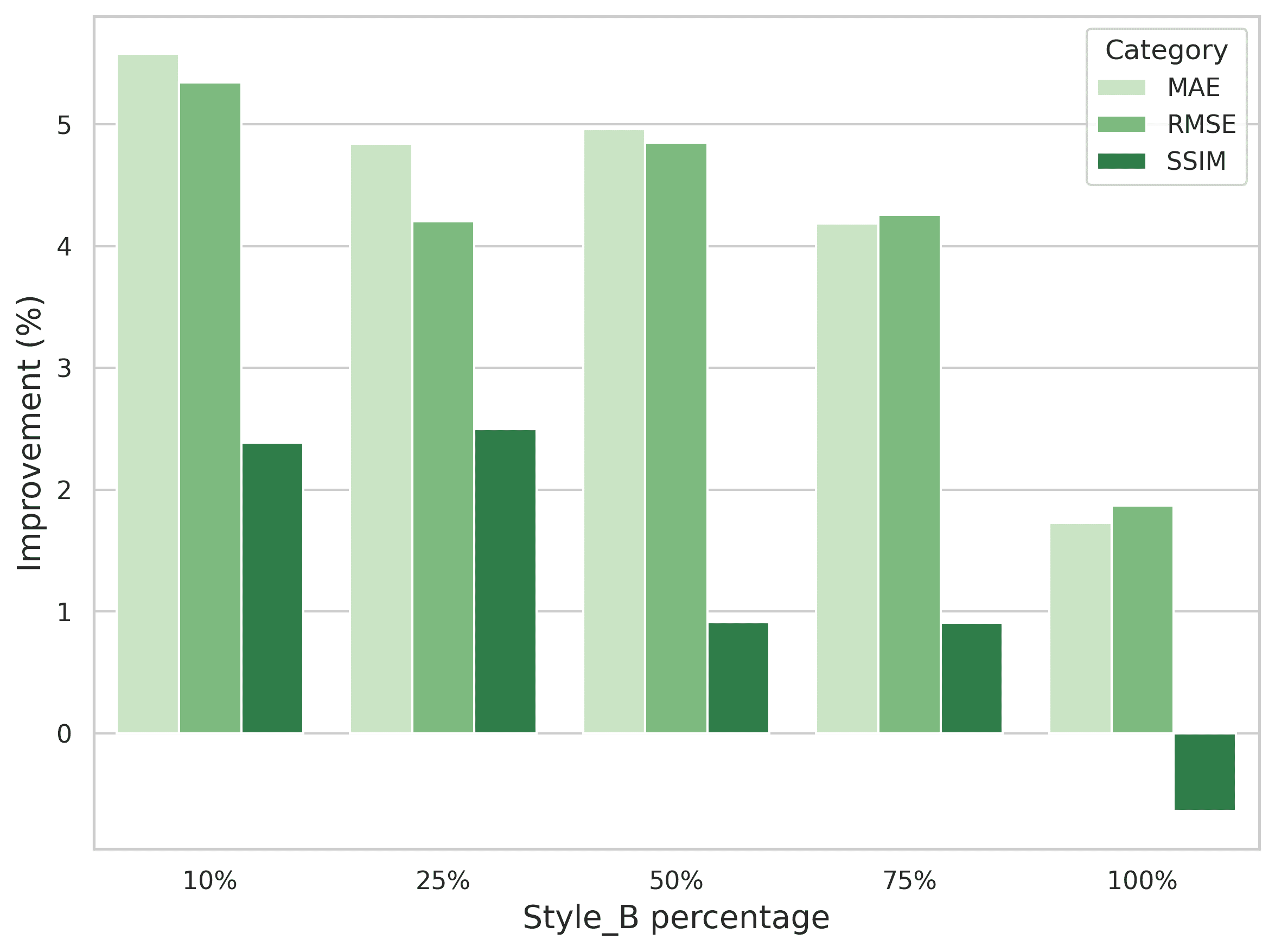}
        \label{fig:SBSA}
    }
    \caption{Bar plot showing generalization improvement of LoRA-PFM over full finetuning. PFM was finetuned with various percentages of the training dataset (Style B) and tested with FlatVel B (blue), CurveFault B (red), and Style A (green).}
    \label{data_per_StyleB}
\end{figure*}

Style B is a relatively complex dataset and when we finetune PFM with Style B and test over FlatVel B, CurveFault B and Style A. Style B contains randomly oriented smooth as well as sharp boundaries. CurveFault B and Style A have many similar features within them and hence we can observe that the improvement of LoRA-PFM with respect to FFT-PFM is relatively small with the increasing size of the dataset. For CurveFault B, we also observe a crossover at 50\% where FFT-PFM outperforms LoRA-PFM, as shown in Fig. \ref{data_per_StyleB}. For Style A, we observe a gradual convergence of the improvement bar as the dataset size increases. The above-discussed improvement can also be visualized by the Fig \ref{fig:ood_dat_per_train}(d). However, FlatVel B is a simple dataset but the features shared by FlatVel B are distinct from Style B. Hence the improvement bar shows LoRA-PFM performs better than FFT-PFM with a small dataset size but as the dataset size increases beyond 50\% the reverse is true. 

The Table \ref{tab:OOD_finetune_FlatVelB}, \ref{tab:OOD_finetune_CurvefaultB}, \ref{tab:OOD_finetune_Style_A}, and \ref{tab:OOD_finetune_Style_B} shows that LoRA-PFM outperform FFT-PFM in all the three metrics.
\begin{figure*}
    \centering
    \includegraphics[width=\linewidth]{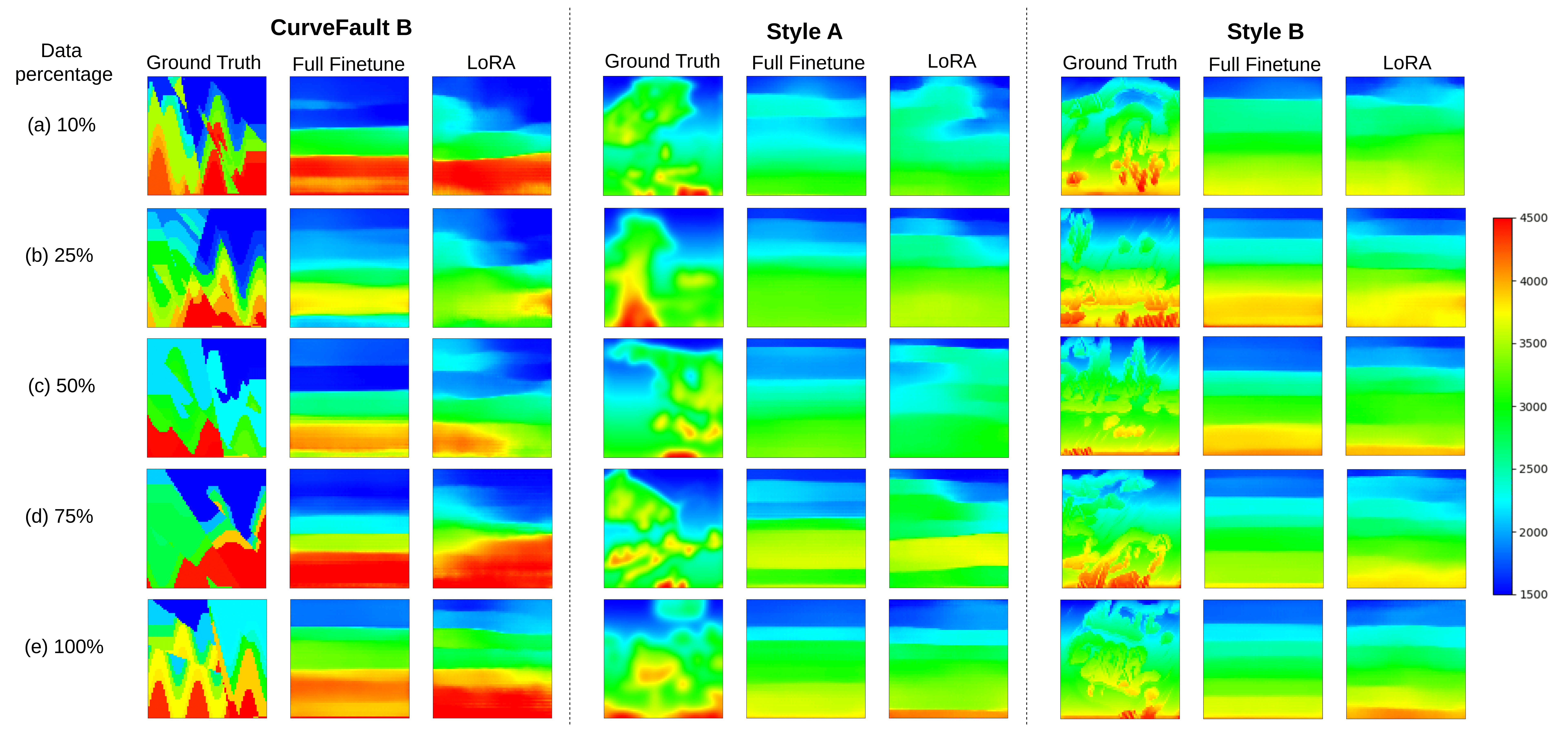}
    \caption{Train: FlatVel B with different training dataset \% and Test: CurveFault B, Style A and Style B}
    \label{fig:ood_dat_per_train_fvb}
\end{figure*}

\begin{figure*}
    \centering
    \includegraphics[width=\linewidth]{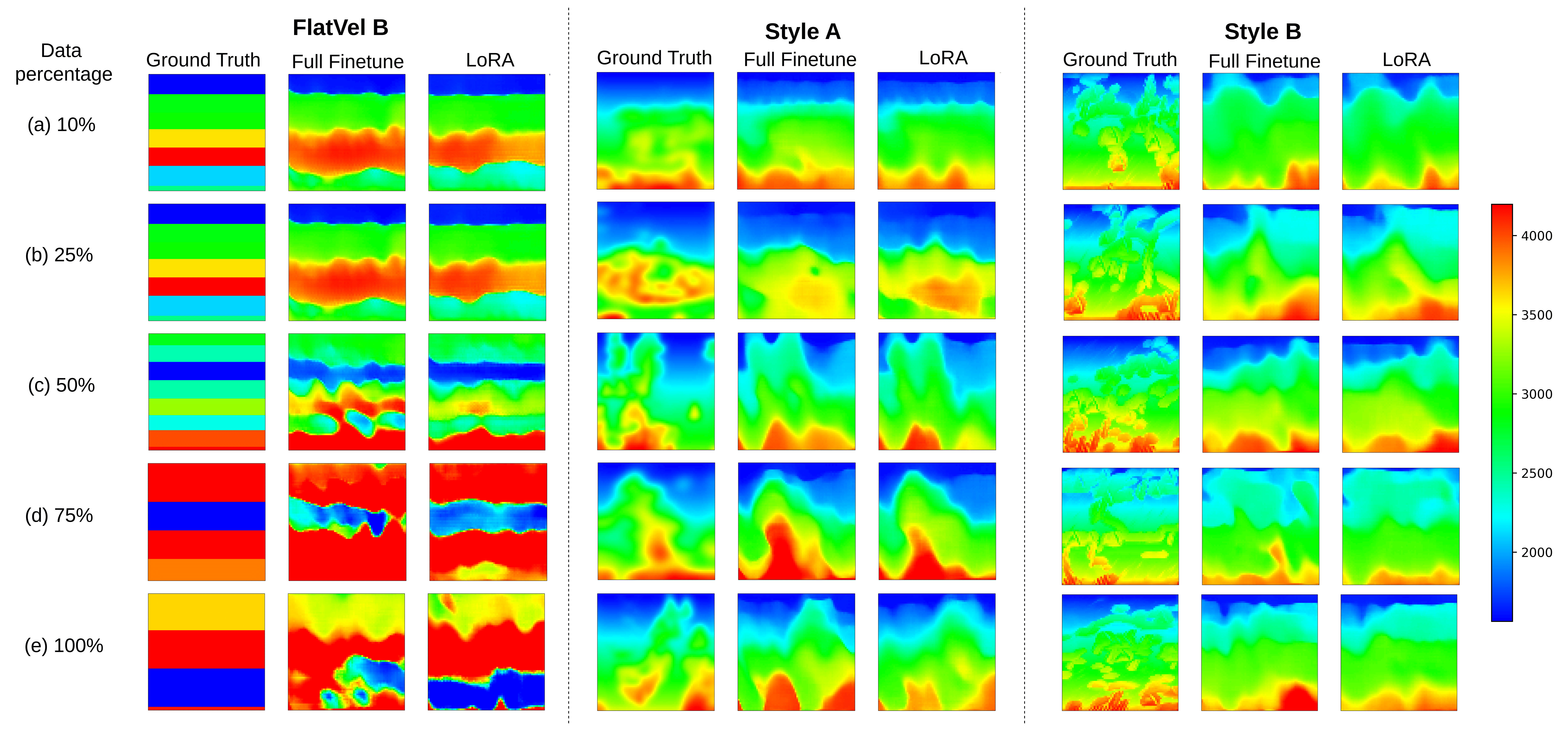}
    \caption{Train: Curvefault B with different training dataset \% and Test: FlatVel B, Style A and Style B}
    \label{fig:ood_dat_per_train_cfb}
\end{figure*}

\begin{figure*}
    \centering
    \includegraphics[width=\linewidth]{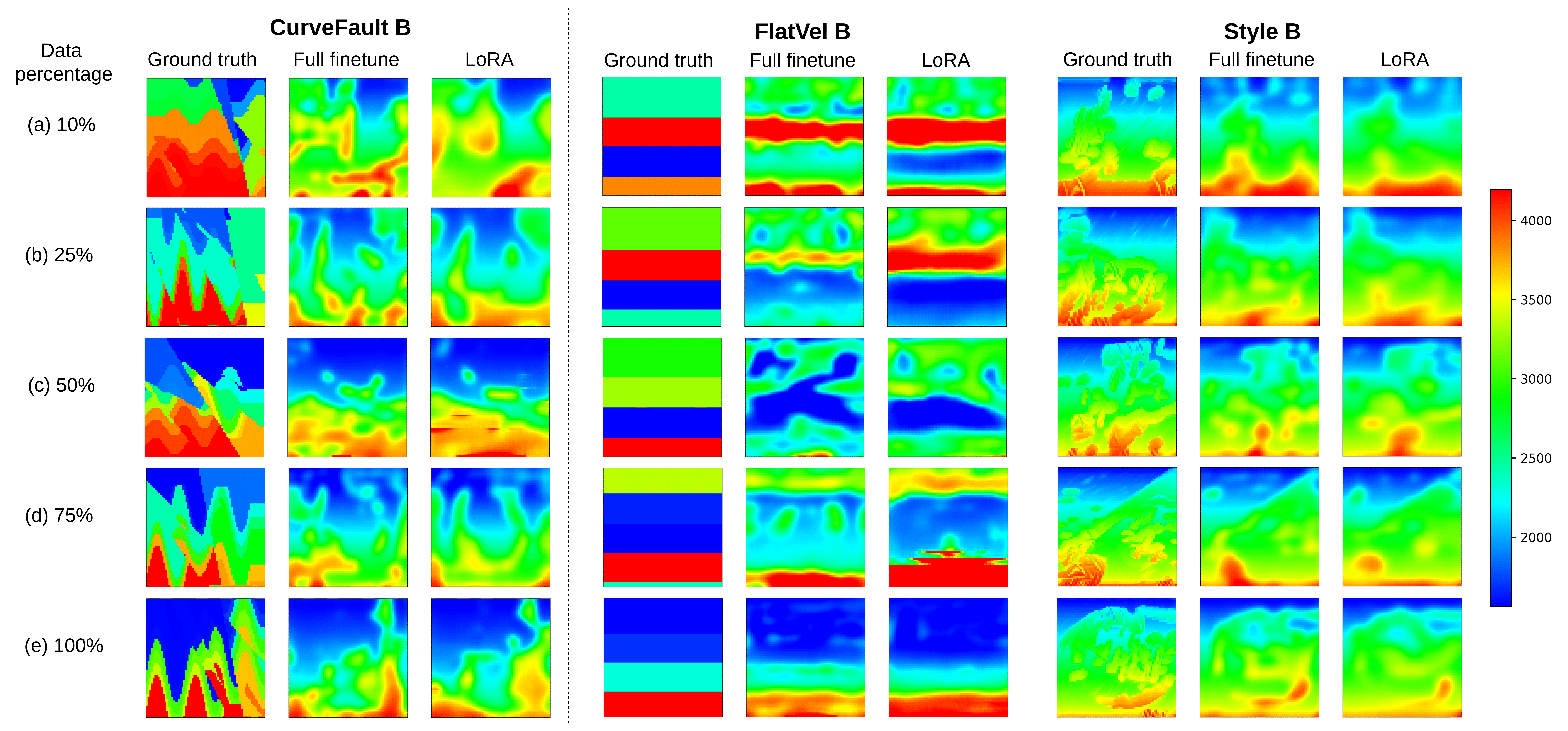}
    \caption{Train: Style A with different training dataset \% and Test: FlatVel B, CurveFault B and Style B}
    \label{fig:ood_dat_per_train_sta}
\end{figure*}

\begin{figure*}
    \centering
    \includegraphics[width=\linewidth]{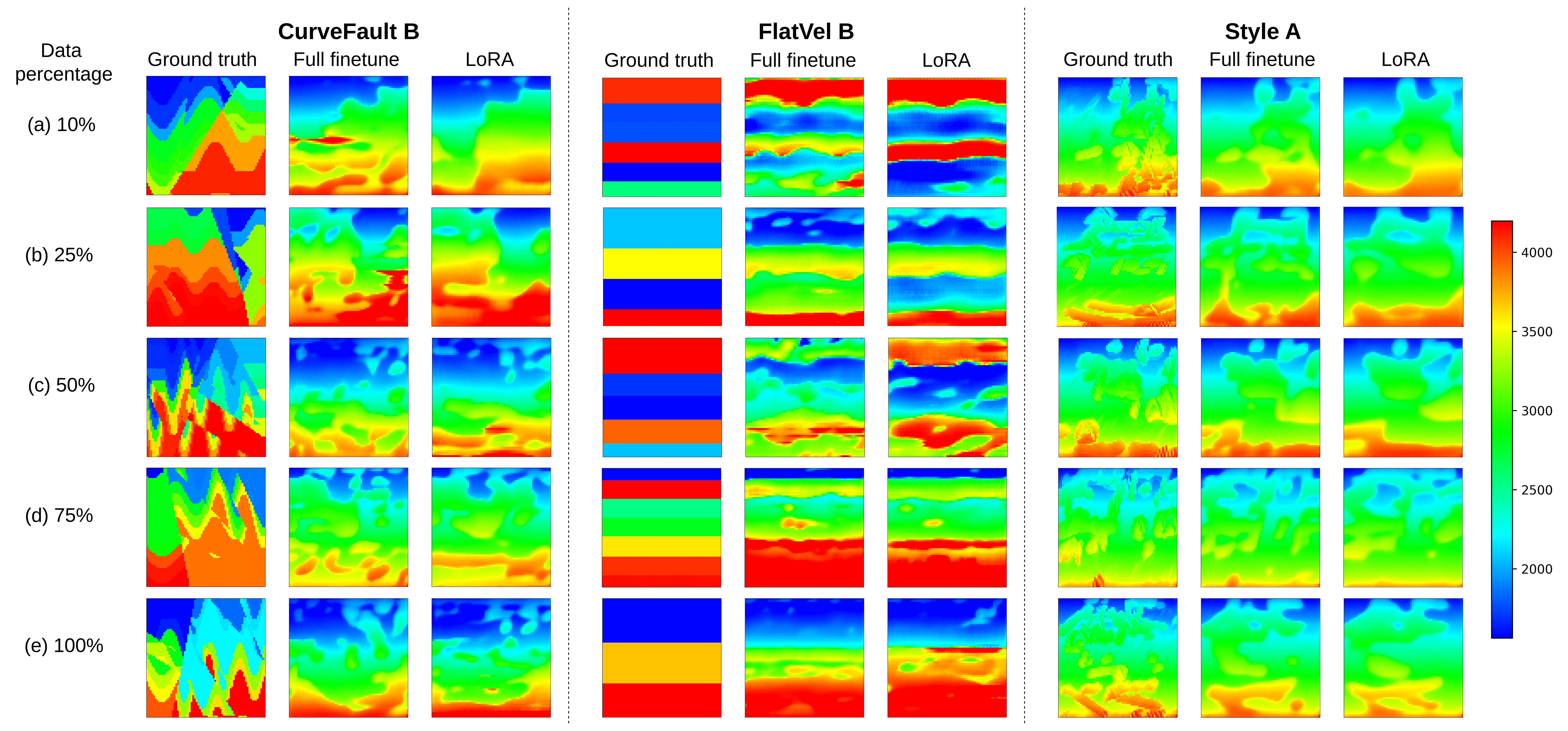}
    \caption{Train: Style B with different training dataset \% and Test: FlatVel B, CurveFault B and Style A}
    \label{fig:ood_dat_per_train_stb}
\end{figure*}


\onecolumn

\begin{table}[!h]
    \centering
    \begin{tabular}{|l|l|l|l|l|l|}
    \hline
    Test Dataset & Train Dataset(\%) & Method & MAE ($\downarrow$) & RMSE ($\downarrow$) & SSIM ($\uparrow$) \\
    \hline
    \multirow{10}{*}{CurveFault B} & \multirow{2}{6em}{10} & FFT-PFM & 0.277 & 0.372 & 0.465 \\ 
    &  & LoRA-PFM & \textbf{0.259} & \textbf{0.353} & \textbf{0.504} \\ 
    \cline{2-6}
    & \multirow{2}{*}{25} & FFT-PFM & 0.277 & 0.372 & 0.465 \\ 
    &  & LoRA-PFM & \textbf{0.259} & \textbf{0.353} & \textbf{0.504} \\
    \cline{2-6}
    & \multirow{2}{*}{50} & FFT-PFM & 0.308 & 0.417 & 0.426 \\ 
    &  & LoRA-PFM & \textbf{0.263} & \textbf{0.363} & \textbf{0.489} \\
    \cline{2-6}
    & \multirow{2}{*}{75} & FFT-PFM & 0.312 & 0.178 & 0.421 \\ 
    &  & LoRA-PFM & \textbf{0.276} & \textbf{0.383} & \textbf{0.470} \\
    \cline{2-6}
    & \multirow{2}{*}{100} & FFT-PFM & 0.321 & 0.434 & 0.420 \\ 
    &  & LoRA-PFM & \textbf{0.271} & \textbf{0.374} & \textbf{0.471} \\
    
    \hline
    
    \multirow{10}{*}{Style A} & \multirow{2}{6em}{10} & FFT-PFM & 0.169 & 0.225 & 0.676 \\ 
    &  & LoRA-PFM & \textbf{0.164} & 0.228 & \textbf{0.696} \\ 
    \cline{2-6}
    & \multirow{2}{*}{25} & FFT-PFM & 0.183 & 0.244 & 0.648 \\ 
    &  & LoRA-PFM & \textbf{0.156} & \textbf{0.213} & \textbf{0.685} \\
    \cline{2-6}
    & \multirow{2}{*}{50} & FFT-PFM & 0.195 & 0.260 & 0.628 \\ 
    &  & LoRA-PFM & \textbf{0.163} & \textbf{0.223} & \textbf{0.672} \\
    \cline{2-6}
    & \multirow{2}{*}{75} & FFT-PFM & 0.188  & 0.252 & 0.648 \\ 
    &  & LoRA-PFM & \textbf{0.167} & \textbf{0.229} & \textbf{0.667} \\
    \cline{2-6}
    & \multirow{2}{*}{100} & FFT-PFM & 0.202 & 0.272 & 0.624 \\ 
    &  & LoRA-PFM & \textbf{0.164} & \textbf{0.223} & \textbf{0.678} \\
    
    \hline
    
    \multirow{10}{*}{Style B} & \multirow{2}{6em}{10} & FFT-PFM & \textbf{0.129} & \textbf{0.163} & 0.578 \\ 
    &  & LoRA-PFM & 0.131 & 0.167 & \textbf{0.590} \\ 
    \cline{2-6}
    & \multirow{2}{*}{25} & FFT-PFM & 0.133 & 0.167 & 0.560 \\ 
    &  & LoRA-PFM & \textbf{0.120} & \textbf{0.151} & \textbf{0.584} \\
    \cline{2-6}
    & \multirow{2}{*}{50} & FFT-PFM & 0.148 & 0.187 & 0.549 \\ 
    &  & LoRA-PFM & \textbf{0.125} & \textbf{0.158} & \textbf{0.578} \\
    \cline{2-6}
    & \multirow{2}{*}{75} & FFT-PFM & 0.144 & 0.181 & 0.559 \\ 
    &  & LoRA-PFM & \textbf{0.126} & \textbf{0.160} & \textbf{0.573} \\
    \cline{2-6}
    & \multirow{2}{*}{100} & FFT-PFM & 0.151 & 0.191 & 0.545 \\ 
    &  & LoRA-PFM & \textbf{0.122} & \textbf{0.154} & \textbf{0.575} \\
    \hline
    \end{tabular}
    \caption{OOD Generalization for Train set FlatVelB and  Test set CurveFaultB, StyleA, StyleB}
    \label{tab:OOD_finetune_FlatVelB}
\end{table}

% *****************************************
% *************CurvefaultB ****************
% *****************************************

\begin{table}[!h]
    \centering
    \begin{tabular}{|l|l|l|l|l|l|}
    \hline
    Test Dataset & Train Dataset(\%) & Method & MAE ($\downarrow$) & RMSE ($\downarrow$) & SSIM ($\uparrow$) \\
    \hline
    \multirow{10}{*}{FlatVel B} & \multirow{2}{6em}{10} & FFT-PFM & 0.243 & 0.374 & 0.640 \\ 
    &  & LoRA-PFM & \textbf{0.156} & \textbf{0.271} & \textbf{0.720} \\ 
    \cline{2-6}
    & \multirow{2}{*}{25} & FFT-PFM & 0.242 & 0.376 & 0.640 \\ 
    &  & LoRA-PFM & \textbf{0.166} & \textbf{0.289} & \textbf{0.711} \\
    \cline{2-6}
    & \multirow{2}{*}{50} & FFT-PFM & 0.238 & 0.379 & 0.629 \\ 
    &  & LoRA-PFM & \textbf{0.214} & \textbf{0.360} & \textbf{0.656} \\
    \cline{2-6}
    & \multirow{2}{*}{75} & FFT-PFM & 0.321 & 0.487 & 0.544 \\ 
    &  & LoRA-PFM & \textbf{0.227} & \textbf{0.383} & \textbf{0.638} \\
    \cline{2-6}
    & \multirow{2}{*}{100} & FFT-PFM & 0.310 & 0.472 & 0.555 \\ 
    &  & LoRA-PFM & \textbf{0.248} & \textbf{0.410} & \textbf{0.619} \\
    
    \hline
    
    \multirow{10}{*}{Style A} & \multirow{2}{6em}{10} & FFT-PFM & 0.143 & 0.196 & 0.720 \\ 
    &  & LoRA-PFM & \textbf{0.128} & \textbf{0.176} & \textbf{0.732} \\ 
    \cline{2-6}
    & \multirow{2}{*}{25} & FFT-PFM & 0.131 & 0.18 & 0.731 \\ 
    &  & LoRA-PFM & \textbf{0.126} & \textbf{0.174} & \textbf{0.736} \\
    \cline{2-6}
    & \multirow{2}{*}{50} & FFT-PFM & 0.139 & 0.196 & 0.730 \\ 
    &  & LoRA-PFM & \textbf{0.125} & \textbf{0.174} & \textbf{0.737} \\
    \cline{2-6}
    & \multirow{2}{*}{75} & FFT-PFM & 0.150 & 0.213 & 0.708 \\ 
    &  & LoRA-PFM & \textbf{0.122} & \textbf{0.171} & \textbf{0.744} \\
    \cline{2-6}
    & \multirow{2}{*}{100} & FFT-PFM & 0.137 & 0.191 & 0.720 \\ 
    &  & LoRA-PFM & \textbf{0.119} & \textbf{0.166} & \textbf{0.743} \\
    
    \hline
    
    \multirow{10}{*}{Style B} & \multirow{2}{6em}{10} & FFT-PFM & 0.114 & 0.148 & 0.600 \\ 
    &  & LoRA-PFM & \textbf{0.108} & \textbf{0.138} & \textbf{0.604} \\ 
    \cline{2-6}
    & \multirow{2}{*}{25} & FFT-PFM & 0.111 & 0.144 & 0.607 \\ 
    &  & LoRA-PFM & \textbf{0.108} & \textbf{0.140} & \textbf{0.609} \\
    \cline{2-6}
    & \multirow{2}{*}{50} & FFT-PFM & 0.115 & 0.148 & 0.604 \\ 
    &  & LoRA-PFM & \textbf{0.107} & \textbf{0.139} & \textbf{0.608} \\
    \cline{2-6}
    & \multirow{2}{*}{75} & FFT-PFM & 0.121 & 0.159 & 0.587 \\ 
    &  & LoRA-PFM & \textbf{0.107} & \textbf{0.138} & \textbf{0.608} \\
    \cline{2-6}
    & \multirow{2}{*}{100} & FFT-PFM & 0.118 & 0.156 &  0.592 \\ 
    &  & LoRA-PFM & \textbf{0.104} & \textbf{0.134} & \textbf{0.610} \\
    \hline
    \end{tabular}
    \caption{OOD Generalization for Train set CurveFaultB and Test set FlatVelB, StyleA, StyleB}
    \label{tab:OOD_finetune_CurvefaultB}
\end{table}

% **************************************
% ************* Style A ****************
% **************************************
\begin{table}[!h]
    \centering
    \begin{tabular}{|l|l|l|l|l|l|}
    \hline
    Test Dataset & Train Dataset(\%) & Method & MAE ($\downarrow$) & RMSE ($\downarrow$) & SSIM ($\uparrow$) \\
    \hline
    \multirow{10}{*}{FlatVel B} & \multirow{2}{6em}{10} & FFT-PFM & 0.322 & 0.440 & 0.486 \\ 
    &  & LoRA-PFM & \textbf{0.236} & \textbf{0.344} & \textbf{0.569} \\ 
    \cline{2-6}
    & \multirow{2}{*}{25} & FFT-PFM & 0.346 & 0.462 & 0.460 \\ 
    &  & LoRA-PFM & \textbf{0.239} & \textbf{0.346} & \textbf{0.575} \\
    \cline{2-6}
    & \multirow{2}{*}{50} & FFT-PFM & 0.365 & 0.485 & 0.440 \\ 
    &  & LoRA-PFM & \textbf{0.248} & \textbf{0.353} & \textbf{0.551} \\
    \cline{2-6}
    & \multirow{2}{*}{75} & FFT-PFM & 0.363 & 0.481 & 0.448 \\ 
    &  & LoRA-PFM & \textbf{0.275} & \textbf{0.408} & \textbf{0.511} \\
    \cline{2-6}
    & \multirow{2}{*}{100} & FFT-PFM & 0.358 & 0.478 & 0.440 \\ 
    &  & LoRA-PFM & \textbf{0.264} & \textbf{0.386} & \textbf{0.511} \\
    
    \hline
    
    \multirow{10}{*}{CurveFault B} & \multirow{2}{6em}{10} & FFT-PFM & 0.300 & 0.403 & 0.435 \\ 
    &  & LoRA-PFM & \textbf{0.272} & \textbf{0.367} & \textbf{0.491} \\ 
    \cline{2-6}
    & \multirow{2}{*}{25} & FFT-PFM & 0.298 & 0.4 & 0.445 \\ 
    &  & LoRA-PFM & \textbf{0.267} & \textbf{0.360} & \textbf{0.495} \\
    \cline{2-6}
    & \multirow{2}{*}{50} & FFT-PFM & 0.293 & 0.396 & 0.446 \\ 
    &  & LoRA-PFM & \textbf{0.264} & \textbf{0.356} & \textbf{0.482} \\
    \cline{2-6}
    & \multirow{2}{*}{75} & FFT-PFM & 0.287 & 0.387 & 0.451 \\ 
    &  & LoRA-PFM & \textbf{0.267} & \textbf{0.363} & \textbf{0.471} \\
    \cline{2-6}
    & \multirow{2}{*}{100} & FFT-PFM & 0.280 & 0.379 & 0.457 \\ 
    &  & LoRA-PFM & \textbf{0.256} & \textbf{0.349} & \textbf{0.474} \\
    
    \hline
    
    \multirow{10}{*}{Style B} & \multirow{2}{6em}{10} & FFT-PFM & 0.099 & 0.137 & 0.639 \\ 
    &  & LoRA-PFM & 0.095 & 0.131 & 0.650 \\ 
    \cline{2-6}
    & \multirow{2}{*}{25} & FFT-PFM & 0.094 & 0.133 & 0.652 \\ 
    &  & LoRA-PFM & \textbf{0.090} & \textbf{0.122} & \textbf{0.658} \\
    \cline{2-6}
    & \multirow{2}{*}{50} & FFT-PFM & 0.092 & 0.130 & \textbf{0.663} \\ 
    &  & LoRA-PFM & 0.092 & 0.130 & 0.660 \\
    \cline{2-6}
    & \multirow{2}{*}{75} & FFT-PFM & 0.091 & 0.126 & \textbf{0.667} \\ 
    &  & LoRA-PFM & \textbf{0.089} & 0.126 & 0.665 \\
    \cline{2-6}
    & \multirow{2}{*}{100} & FFT-PFM & \textbf{0.085} & 0.118 & \textbf{0.680} \\ 
    &  & LoRA-PFM & 0.088 & 0.118 & 0.667 \\
    \hline
    \end{tabular}
    \caption{OOD Generalization for Train set StyleA and Test set FlatVelB, CurveFaultB, StyleB}
    \label{tab:OOD_finetune_Style_A}
\end{table}

% **************************************
% ************* Style B ****************
% **************************************
\begin{table}[!h]
    \centering
    \begin{tabular}{|l|l|l|l|l|l|}
    \hline
    Test Dataset & Train Dataset(\%) & Method & MAE ($\downarrow$) & RMSE ($\downarrow$) & SSIM ($\uparrow$) \\
    \hline
    \multirow{10}{*}{FlatVel B} & \multirow{2}{6em}{10} & FFT-PFM & 0.319 & 0.442 & 0.482 \\ 
    &  & LoRA-PFM & \textbf{0.270} & \textbf{0.392} & \textbf{0.555} \\ 
    \cline{2-6}
    & \multirow{2}{*}{25} & FFT-PFM & 0.383 & 0.504 & 0.397 \\ 
    &  & LoRA-PFM & \textbf{0.276} & \textbf{0.382} & \textbf{0.513} \\
    \cline{2-6}
    & \multirow{2}{*}{50} & FFT-PFM & 0.487 & 0.614 & 0.348 \\ 
    &  & LoRA-PFM & \textbf{0.329} & \textbf{0.450} & \textbf{0.431} \\
    \cline{2-6}
    & \multirow{2}{*}{75} & FFT-PFM & 0.504 & 0.636 & 0.334 \\ 
    &  & LoRA-PFM & \textbf{0.409} & \textbf{0.569} & \textbf{0.357} \\
    \cline{2-6}
    & \multirow{2}{*}{100} & FFT-PFM & 0.423 & 0.542 & 0.370 \\ 
    &  & LoRA-PFM & \textbf{0.402} & \textbf{0.550} & 0.370 \\
    
    \hline
    
    \multirow{10}{*}{CurveFault B} & \multirow{2}{6em}{10} & FFT-PFM & 0.267 & 0.360 & 0.428 \\ 
    &  & LoRA-PFM & \textbf{0.247} & \textbf{0.331} & \textbf{0.461} \\ 
    \cline{2-6}
    & \multirow{2}{*}{25} & FFT-PFM & 0.263 & 0.349 & 0.419 \\ 
    &  & LoRA-PFM & \textbf{0.242} & \textbf{0.320} & \textbf{0.469} \\
    \cline{2-6}
    & \multirow{2}{*}{50} & FFT-PFM & 0.271 & 0.357 & 0.427 \\ 
    &  & LoRA-PFM & \textbf{0.253} & \textbf{0.340} & \textbf{0.434} \\
    \cline{2-6}
    & \multirow{2}{*}{75} & FFT-PFM & 0.267 & 0.352 & \textbf{0.429} \\ 
    &  & LoRA-PFM & \textbf{0.276} & \textbf{0.392} & 0.401 \\
    \cline{2-6}
    & \multirow{2}{*}{100} & FFT-PFM & \textbf{0.255} & \textbf{0.339} & \textbf{0.452} \\ 
    &  & LoRA-PFM & 0.276 & 0.391 & 0.410 \\
    
    \hline
    
    \multirow{10}{*}{Style A} & \multirow{2}{6em}{10} & FFT-PFM & 0.129 & 0.173 & 0.746 \\ 
    &  & LoRA-PFM & \textbf{0.122} & \textbf{0.164} & \textbf{0.764} \\ 
    \cline{2-6}
    & \multirow{2}{*}{25} & FFT-PFM & 0.127 & 0.170 & 0.751 \\ 
    &  & LoRA-PFM & \textbf{0.121} & \textbf{0.163} & \textbf{0.770} \\
    \cline{2-6}
    & \multirow{2}{*}{50} & FFT-PFM & 0.124 & 0.169 & 0.764 \\ 
    &  & LoRA-PFM & \textbf{0.118} & \textbf{0.161} & \textbf{0.771} \\
    \cline{2-6}
    & \multirow{2}{*}{75} & FFT-PFM & 0.122 & 0.168 & 0.769 \\ 
    &  & LoRA-PFM & \textbf{0.117} & \textbf{0.161} & \textbf{0.776} \\
    \cline{2-6}
    & \multirow{2}{*}{100} & FFT-PFM & 0.117 & 0.162 & \textbf{0.785} \\ 
    &  & LoRA-PFM & \textbf{0.115} & \textbf{0.159} & 0.780 \\
    \hline
    \end{tabular}
    \caption{OOD Generalization for Train set StyleB and Test set FlatVelB, CurveFaultB, StyleA}
    \label{tab:OOD_finetune_Style_B}
\end{table}

% \end{document}
\end{document}